\documentclass[11pt]{article}
\usepackage[margin=1in]{geometry}
\usepackage[T1]{fontenc}
\usepackage[utf8]{inputenc}
\usepackage{lmodern,setspace}
\usepackage{amsmath,amssymb,amsthm}
\usepackage{graphicx,booktabs,longtable,array,threeparttable}
\usepackage[round,authoryear]{natbib}
\usepackage[colorlinks=true,allcolors=black]{hyperref}
\usepackage[font=small,labelfont=normalfont,textfont=normalfont,labelsep=period,justification=raggedright,singlelinecheck=false]{caption}
\usepackage{placeins}
\hypersetup{hypertexnames=false}
\renewenvironment{abstract}{\begin{center}\normalfont Abstract\end{center}\begin{quote}\small}{\end{quote}}
\newcommand{\tabnote}[1]{\par\smallskip\begin{minipage}{\linewidth}\small #1\end{minipage}}

\title{AI and the research assistant:\\ Evidence from economics working papers\thanks{I thank Lukas Althoff, Joshua Gans, Alex Imas, Benjamin Jones, and Preston McAfee for feedback on the paper. I also acknowledge feedback provided by refine.ink. Anthropic's Claude Fable coordinated the revision and edited the text, Google's Gemini 4 Argon drafted the text, and OpenAI's GPT-6 Astra assisted with code preparation and documentary checks. Appendix A describes the text-coding procedures. Cite this paper as: Fourie, Johan. 2026. ``AI and the research assistant: Evidence from economics working papers.'' Working Paper, Department of Economics, Stellenbosch University.}}
\author{Johan Fourie\thanks{Department of Economics, Stellenbosch University. Email: \href{mailto:johanf@sun.ac.za}{johanf@sun.ac.za}.}}
\date{}
\begin{document}\maketitle
\begin{abstract}
Studies of AI in science observe isolated tasks or the authors on a byline. They do not observe the research assistants whose data cleaning, coding, literature search, and replication work language models now perform. Economics working papers record this junior labour in their acknowledgements, next to the full text. Using 24 working-paper series through October 2026, I compare papers by the same economist that do and do not disclose AI use, alongside a detector of AI-like prose. Disclosing papers credit 0.548 more named assistants than other papers by the same economist, two-thirds of the mean, and run longer, mostly in their appendices. Disclosures of operational use, such as coding and data work, carry the whole assistant difference; writing-only disclosures and AI-like prose show no consistent association with assistants or pages. The paper brings acknowledged junior labour and selective team disclosure into the study of scientific production. Within an economist's portfolio, operational AI accompanies wider, assistant-supported projects.
\end{abstract}
\vspace{1em}\noindent\textbf{Keywords:} artificial intelligence; research assistants; scientific teams; voluntary disclosure; economics profession

\vspace{0.25em}

\vspace{0.5em}
\newpage
\section{Introduction}\label{introduction}

Before an economics paper reaches a working-paper series, someone cleans the raw files, writes the estimation code, searches the literature, and assembles the appendices. That work has long belonged to the research assistant. Large language models now do much of it in minutes \citep{korinek2023,korinek2025}. Payroll and résumé studies find lower entry-level hiring in occupations exposed to AI \citep{brynjolfsson2026canaries,hosseini2026seniority}, and \citet{garicanorayo2025} show how automation can erode the apprenticeship that pays for a junior's training. Studies of AI in science, however, examine isolated tasks or the authors on a byline \citep{noy2023,kusumegi2025,wang2026}. None observes the research assistants credited behind a paper.

I ask whether generative AI substitutes for or complements the research assistant in economics working papers. Working papers record assistance in their acknowledgements, next to the byline, and preserve the full text. Across 56,585 records in 24 series through October 2026, 23 institutional series and the economics arXiv, I link assistance credits to two AI indicators: an explicit statement that the authors used AI, and a writing flag from a detector of AI-like economics prose \citep{detector2026,jabarianimas2025}. The main regressions compare post-2022 papers written by the same recurring researcher, with researcher, source-quarter, and field fixed effects. Comparing papers within one researcher's portfolio removes persistent differences in funding and in the habit of thanking people. The credits measure contributions to a paper, not hours or employment.

Within an economist's portfolio, papers that disclose AI use credit more research assistants and run longer. Disclosing papers credit 0.548 more named assistants, two-thirds of the mean of 0.83 on papers without a disclosure, and they run about a tenth longer. Both differences survive the adjustment for the registered family of twenty tests. Disclosing papers are also more likely to credit any assistant and carry slightly more authors, but these differences do not survive the adjustment. The disclosure variation sits in a few effective sources, and the assistant estimates move with how contributions are documented and how sources are weighted. Section 5 reports those sensitivities.

The writing flag, by contrast, shows no consistent association with authors, named assistants, or pages. The two indicators also barely overlap: few disclosing papers carry the flag, and few flagged papers disclose. A simple project-choice framework explains why the indicators diverge. Cheaper execution lowers the hours a fixed study needs, but it raises the demand for assistance when researchers widen the study. A saving on writing that is separable from project design leaves inputs unchanged.

Two exploratory splits test that mechanism. Four in five assessable disclosures report coding, data, analysis, or derivation rather than writing alone. Within researcher, those operational disclosures carry the whole assistant difference, 0.679 more named assistants, whereas writing-only statements show none. Operational disclosures also account for the extra length, and most of that length sits in appendices rather than in the main text. The aggregate series tell the same story: in the institutional series the share of papers crediting any assistant stays between 27\% and 32\% from 2022 to 2026 while AI disclosure rises from zero to about one paper in six. An exploratory check on multiple-testing corrections, a possible fingerprint of AI in the analysis, proves too rare and noisy to serve as a third indicator.

Voluntary disclosure remains the main threat to this reading. An unattributed team statement dilutes personal stigma across coauthors \citep{reif2025}, so the model predicts more authors on disclosing papers without any change in production. But it predicts nothing for assistants or pages. In the data, solo papers disclose more often than coauthored papers in the raw comparison, but that excess vanishes within researcher, and the data reject the account in which each author independently generates a disclosure.

The paper makes two contributions. First, it brings acknowledged junior labour into the economics of generative AI and scientific production. \citet{brynjolfsson2026canaries} and \citet{hosseini2026seniority} document lower junior hiring in exposed jobs, whereas \citet{humlum2026} find small average labour-market effects so far. In economics research, operational AI accompanies wider scope and more credited assistants \citep{acemoglu2019,filimonovic2025genai}. Closest in setting, \citet{jiang2026equalizing} find across economists that AI narrowed the productivity gap between researchers with and without research assistants. Within the same economist's portfolio, disclosing projects credit more assistants. Substitution for an absent assistant and expansion with a present one are compatible, and the two studies observe different margins of the same technology. Steady institutional assistant shares and the operational association are consistent with a junior position that persists on paper, which moves the apprenticeship question in \citet{garicanorayo2025} and \citet{wang2026predoc} towards the content of the tasks.

Second, the paper separates operational disclosures from prose resemblance in one corpus and treats the disclosure as a collective decision. \citet{kusumegi2025} and \citet{wang2026} link detected AI prose to output and to smaller author teams; \citet{hosseinioun2026} find larger teams in AI grant proposals. Estimating both indicators together shows that the writing flag yields intervals that include zero, so modest reductions of the kind Wang et al.~report are not ruled out, whereas operational disclosures carry the assistant difference. Observing assistants beside an author difference that is not robust separates delegated execution from coauthorship in theories of knowledge hierarchies and team size \citep{jones2021,garicano2000,ide2025,fourie2026team}. Modelling self-reported AI use under shared attribution connects the measurement of AI in science to voluntary and group disclosure \citep{dye1985disclosure,levy2007committees,onuchicray2023,onuchicramos2025}.

The rest of the paper is structured as follows. Section 2 sets out the project-choice framework that links execution costs and separable writing to project scope and assistant hours. Section 3 describes the 24 series, the assistance and AI measures, the adoption trends, and the multiple-testing check. Section 4 presents the within-researcher specification and the attribution model. Section 5 reports the main estimates, the task-mix and page decompositions, and the sensitivity checks. Section 6 turns to annual researcher portfolios and the training role of research assistance, and Section 7 concludes.

\section{Project choice, assistant hours, and separable writing}\label{sec:framework}

A research team chooses both the authors on a paper and the breadth of the investigation they carry out. Let \(n\geq 1\) denote the number of byline authors and \(x\geq 0\) the project's scope: the extent of empirical and analytical work undertaken. In practice, wider scope means assembling extra datasets, running further specifications, checking edge cases, or building fuller appendices. Combining expertise raises a project's value. Execution, verification, coordination among coauthors, and writing consume resources: \begin{equation}
 U(n,x)=B(n)\log(1+x)-c_o(a_o)x-\frac{jx^2}{2n}-K(n)-C_w(a_w).
 \label{eq:project}
\end{equation} Here \(B(n)\) is the benefit of coauthor expertise and \(c_o(a_o)\) is the unit execution cost, the resources spent per unit of routine operational work under AI capability \(a_o\), with \(c_o'(a_o)<0\). Judgement costs \(jx^2/(2n)\), with \(j>0\), rise with scope and are shared across \(n\) authors. \(K(n)\) captures coordination and opportunity costs, and \(C_w(a_w)\) is the cost of writing.

The team chooses \(n\) and \(x\) to maximise \(U(n,x)\) and undertakes the paper if the surplus clears its outside option (Appendix C). At a fixed author count, stronger operational AI lowers \(c_o(a_o)\) and raises the chosen scope \(x^*(n)\), because extra analyses cost less to run. Cheaper execution thus changes what researchers investigate, much as lower production costs widen firms' product lines \citep{babina2024} and new tasks offset automation \citep{acemoglu2019}. A broader study also raises the marginal value of expertise and spreads the judgement work across coauthors \citep{garicano2000,ide2025}. Scope and team size therefore move together, although an integer roster changes only at thresholds and often stays fixed.

Research assistants enter on the delegated execution margin behind the byline. Suppose each unit of scope takes \(e(a_o)\) hours of routine assistant work, with \(e'(a_o)<0\), so that delegated assistant hours on the project are \(h=e(a_o)x\). Within a smooth choice region for scope, \[
\frac{dh}{da_o}=e'(a_o)x+e(a_o)\frac{dx}{da_o}.
\] The first term is the direct saving on existing tasks: an assistant with a language model cleans data or writes code faster. The second term is the new work created when cheaper execution leads the team to widen the study. Either can dominate.

Hours therefore rise only when the study expands proportionally more than the hours per unit of scope fall. Even the sign of hours does not pin down the number of assistants named on the finished paper. One continuing assistant can absorb the expanded work in more hours, leaving the headcount unchanged. A wider investigation that adds a data source or an estimation step can instead bring a new person into the acknowledgements.

Writing assistance \(a_w\) behaves differently when the writing cost \(C_w(a_w)\) is separable, meaning that it enters additively and depends on neither scope nor author count. Under separability the writing cost cancels from every comparison between candidate designs, as when an author edits or translates the prose after the empirical work is complete. Lower writing costs may help a marginal project clear the outside option, but conditional on production they leave the chosen scope and the human inputs unchanged. Separability fails if drafting shapes the argument itself, as \citet{oatley2008} propose, or if writing effort grows with scope and coordination. Polishing the prose of a settled result is separable in this sense; delegating the first formulation of the argument is not \citep{fourie2026writes}. The mechanism in one sentence: operational AI cuts assistant hours on a fixed study yet can raise the demand for assistance by expanding the study, whereas separable writing help leaves staffing unchanged.

Working papers do not display assistant hours or AI capability directly. AI enters the data only when the authors report using a tool, and human assistance enters only when the acknowledgements credit the work. Authors may disclose a tool when it performed a substantive operation or an outlet asked for a statement, and stay silent when reputational concerns outweigh their reporting norms. Teams that keep fuller records or write longer acknowledgements can also document both AI use and assistant names more often at unchanged staffing. Recorded associations therefore combine project design with reporting choices. That is why Section 5 conditions on acknowledgement documentation and Section 4 models the team's decision to disclose.

\section{Working papers, acknowledgements, and AI indicators}\label{sec:data}

\subsection{Sources}\label{sources}

I assemble working papers from 24 series: 23 institutional series and the economics arXiv collection. University departments, central banks, research networks, and policy institutions across North America, Europe, Japan, and Australia supply the institutional outlets. Public catalogues were accessed on 1 November 2026 and cover January 2015 to 31 October 2026, with arXiv entering in September 2017. To prevent selective retrieval, each series had to reach at least 98\% public acquisition overall and in each year from 2023 to 2026. Eight series fell short and are excluded: Central Bank of Chile, Pretoria, Chicago BFI, Paris School of Economics, Tokyo CIRJE, Toronto, Yale Cowles, and Stanford SIEPR.

Bank of England and Philadelphia Fed, each missing a single 2026 paper, are retained as the two exceptions. Across the retained series, the source frame contains 56,585 selected paper-family records; Appendix A lists every series. File hashes and corroborated title and author information group versions and cross-postings into practical paper families without guaranteeing complete deduplication. Unavailable PDFs and unresolved text extractions remain missing on the affected measures rather than receiving imputed zeros.

Archived working-paper PDFs can be overwritten by later revisions. Cohort trends and the post-2022 regressions therefore use timing-supported papers, a subset that combines independently version-dated files with documents whose printed dates match their issue cohorts, under the assumption that no unrecorded revision altered the measured text. Papers with conflicting dates or later-revision notices stay outside this subset, and I examine the stricter independently dated subset separately.

\subsection{Credited assistance, authors, and pages}\label{credited-assistance-authors-and-pages}

Acknowledgements supply two measures of junior labour: whether a paper explicitly credits any research assistance, and the count of named assistants. Named counts record people whose research-assistance role the passage supports, which separates assistants from commenters, advisers, seminar participants, and administrative staff. Collective credits that thank unnamed assistants as a group count as `any' assistance but receive no invented number, so the named count is missing where names are unresolved. AI tools thanked or described as assistants are excluded from the human counts. Recording zero assistance requires readable acknowledgement text rather than an empty or failed extraction. Both outcomes measure credited contributions to a paper, not hours, pay, headcount across papers, or employment.

Author counts come from the printed byline. Pairing the byline with the acknowledgements separates coauthorship from delegated work: a solo author can draw on several assistants, and a large author team can credit none. Page counts record all pages in the retrieved PDF, including references and appendices. Length reflects how much analysis and supplementary material a paper presents, although layout density varies across series and pages do not measure quality. Source-bound definitions keep these measures comparable across outlets without assuming equal effort per named person.

\subsection{Two AI indicators}\label{two-ai-indicators}

Two indicators capture distinct ways in which generative AI becomes visible in a working paper. The writing flag applies the released strict \texttt{econ-\allowbreak ai-\allowbreak detector} \citep{detector2026} to a consistently selected introduction and body section. The detector is a classifier trained to separate pre-ChatGPT economics articles from language-model text, combining logistic and neural thresholds. \citet{jabarianimas2025} show that detector error rates depend on the threshold and on the text being scored. The strict flag therefore measures resemblance to AI-generated economics prose, not verified use. Short or non-English texts are left missing.

The disclosure indicator records an explicit statement that the authors used generative AI in producing the paper. Candidate passages are retrieved from the full text and classified in context against literal source evidence \citep{ludwig2025}. The coding separates own use in producing the paper from studying AI as a topic, applying conventional tools, leaving attribution ambiguous, or stating non-use. Every positive record, and every candidate surfaced by a corpus-wide audit, received a second review against whole-document passages (Appendix A). No confirmed statement in assessable text forms the reference category, which can contain both undisclosed use and missed statements.

Disclosures also say what the tool did. A disclosure is writing-only when every supported task is drafting, editing, translation, or typesetting, and operational or mixed when any passage reports coding, data preparation, search, analysis, or derivation. About four in five assessable disclosures are operational or mixed, and most of the rest are writing-only. Most economists who disclose AI are reporting what it did to their data or their code, not to their prose.

Post-2022 papers observed on both measures show how weakly prose resemblance and explicit disclosure overlap. Only about one in seven disclosing papers carries the writing flag, and about one in eight flagged papers discloses AI use. The two occur together only modestly more often than they would by chance. Data cleaning or coding leaves no prose trace in an introduction, and light editing need not prompt a disclosure. The regressions therefore enter the writing flag and the disclosure together rather than combining them into one adoption index.

\subsection{Adoption since 2022}\label{adoption-since-2022}

\begin{figure}[htbp]\centering
\includegraphics[width=\linewidth]{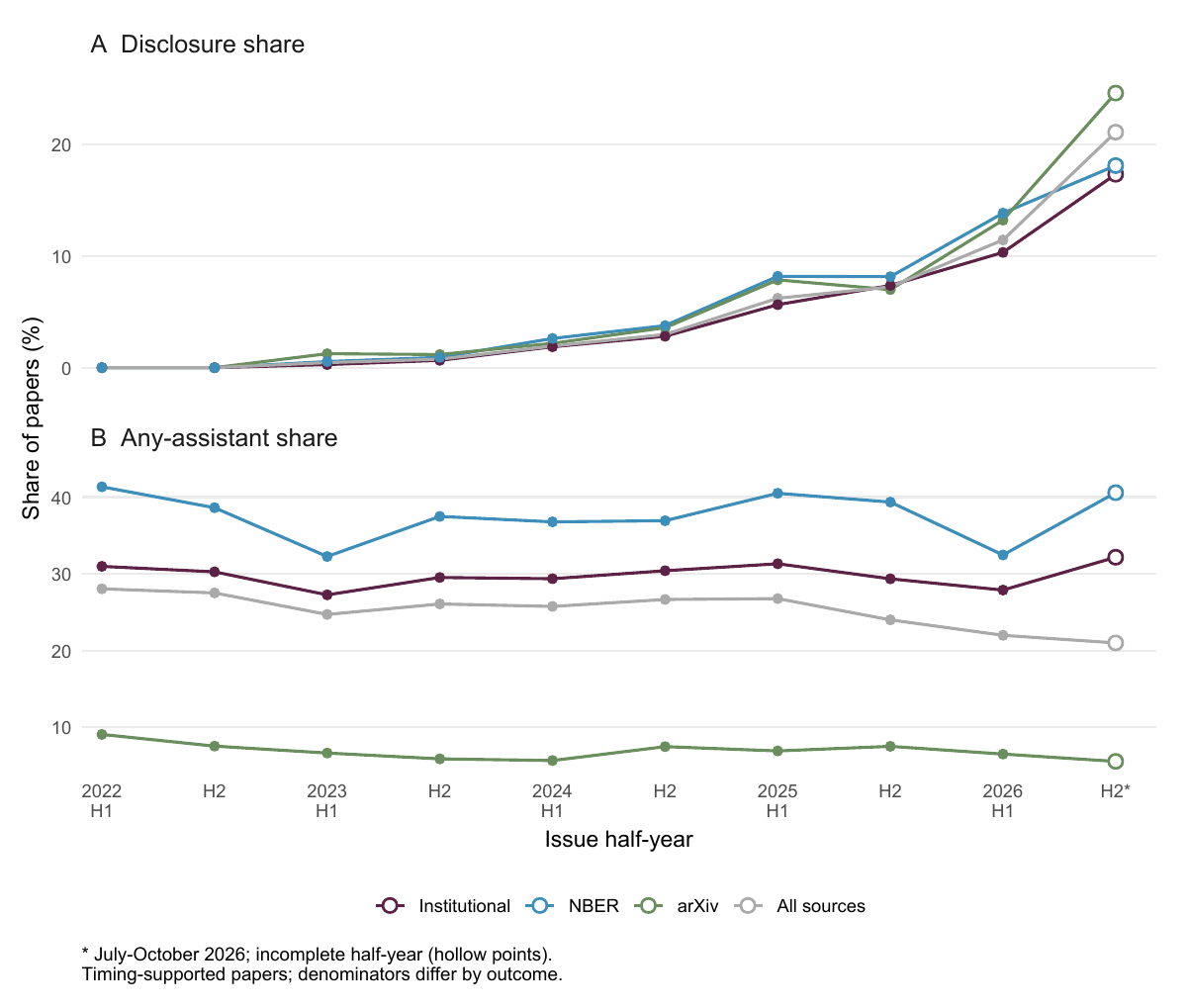}
\caption{AI disclosure and credited assistance}\label{fig:adoption}
\tabnote{Half-year cohorts of timing-supported papers. Panel A shows the disclosure share; Panel B shows the share crediting any assistant. The last point is July--October 2026, an incomplete cohort. Denominators differ by outcome; Appendix~\ref{app:data} reports shares and counts.}\end{figure}

Figure~\ref{fig:adoption} tracks half-year cohorts of timing-supported papers since 2022. It shows the institutional share of papers crediting an assistant holding between 27\% and 32\% while AI disclosure rises from zero to about one paper in six. Across all sources, disclosed AI use climbs from almost nothing in early 2023 to about one paper in five in the incomplete July--October 2026 cohort; Appendix A reports the full table.

Rapid disclosure has not coincided with a drop in acknowledged assistance inside the institutional series, or at NBER. The pooled share across all sources does drift down, but outlet composition explains that decline. ArXiv papers rarely credit assistants, and arXiv's share of the sample grows roughly threefold over the period.

\subsection{Multiple-testing corrections as a possible third indicator}\label{multiple-testing-corrections-as-a-possible-third-indicator}

If language models suggest multiple-testing adjustments when they help with an analysis, those adjustments may mark AI use that no prose detector can see. I search the timing-supported full texts, bibliographies removed, for such corrections. Only a small minority of post-2022 papers apply any correction, and the Holm procedure appears in well under one paper in a hundred. Figure~\ref{fig:mht}, in Appendix B, plots the half-year shares since 2015. The procedures a model defaults to (Holm, Bonferroni, or Benjamini--Hochberg) sit slightly above their pre-2022 trend, whereas the procedures native to economics (Romano--Wolf, List--Shaikh--Xu, Westfall--Young, or Anderson) sit slightly below theirs. Holm relates to disclosure across papers, but the association vanishes within researcher, where very few economists vary in its use. Holm is too rare and too noisy to serve as a third indicator, so the corrections stay an exploratory check in Appendix B. This paper discloses its own AI use and reports Romano--Wolf beside its registered Holm adjustment.

\section{Within-researcher comparison and the disclosure threat}\label{sec:strategy}

\subsection{Specification}\label{specification}

Economists differ persistently in grant funding, prose style, and the habit of thanking junior staff. A cross-sectional comparison of papers with and without AI indicators would mix those personal traits with project-level differences. I therefore compare papers within the same recurring economist's portfolio, following organisational designs that absorb stable individual practices \citep{bertrand2003}. Fixed effects, separate intercepts for each researcher, each outlet-by-quarter cell, and each broad JEL field, hold constant an author's baseline staffing together with outlet cohort norms and field conventions. What remains is the variation in assistant credits, disclosure, and the writing flag across distinct papers written by the same scholar.

For paper \(p\), linked researchers \(i\in\mathcal A_p\), working-paper source \(s\), issue quarter \(q\), and broad field \(f\), I estimate \begin{equation}
 Y_p=\alpha_i+\lambda_{sq}+\mu_f+\beta_W W_p+\beta_D D_p+\theta_1\log L_p+\theta_2\log K_p+\varepsilon_{pi}
 \label{eq:empirical}
\end{equation} on timing-supported papers issued from January 2023 to October 2026. Here \(Y_p\) is a paper outcome, \(W_p\) is the strict writing flag, and \(D_p\) is an explicit disclosure of generative AI use. Both indicators enter together rather than as a combined adoption index. The terms \(\log L_p\) and \(\log K_p\) record selected prose words and scored text windows, which hold constant the detector's opportunity to register a signal. Each paper enters once per included linked researcher with row weight \(1/|\mathcal A_p|\), so every paper has a total weight of one.

Requiring usable text, linked identities, and at least two post-2022 papers per researcher narrows the timing-supported records considerably. After the iterative pruning of singletons, the assistant models retain 7,243 papers by 4,604 recurring researchers, 468 of them disclosing AI use; the byline and length models retain 8,502 papers by 5,285 researchers, 544 of them disclosing. Appendix A traces the sample from the 56,585-record frame, and Appendix B reports pooled models without researcher fixed effects on identical samples.

Standard errors cluster by researcher and by paper, which lets residuals correlate across a scholar's portfolio and across the coauthor rows of one manuscript. But the residual disclosure variation sits in the equivalent of about 3.3 equally informative sources, and shared coauthors link papers across series without sharing a researcher intercept. Source-quarter intercepts do not absorb source-wide shifts in disclosure norms or cross-outlet coauthor covariance; Appendix B.1 evaluates both. Across the registered family of twenty tests, two indicators across ten outcomes, I lead with Romano--Wolf step-down p-values from a researcher-level wild cluster bootstrap and report the registered Holm adjustment beside them. The Romano--Wolf procedure asks how often, across many re-drawn datasets in which no indicator matters, the largest test statistic would be as large as the one observed. It allows for the correlation among the twenty tests that Holm ignores.

The coefficients \(\beta_W\) and \(\beta_D\) measure conditional associations across projects within an economist's recent output. Neither estimates a before-and-after change on a fixed investigation, and a scholar's first flagged or disclosing paper is never treated as an adoption date. Defining treatment timing from an observed output feature would induce mechanical pre-trends, because prolific authors cross any threshold earlier \citep{renault2026}. Within a portfolio, an economist can deploy AI on an ambitious empirical study, omit it on a short note, or report it selectively. Separating scope expansion from selective reporting requires a model of why a team chooses to disclose.

\subsection{Disclosure as a decision}\label{sec:attribution-model}

A detector score is computed from the text. A disclosure is a public statement chosen by the authors after the work is done. \citet{reif2025} document that observers penalise workers who use AI assistance, which gives scholars a reputational reason to hesitate before reporting a tool. On a solo paper, an acknowledgement of AI use identifies the one author as the operator. On a coauthored paper, an unattributed statement reveals that someone on the team used AI without naming who ran the model. Coauthorship can therefore dilute personal attribution even when the technology changes no team. This connects self-reporting in science to voluntary and committee disclosure \citep{dye1985disclosure,levy2007committees,onuchicray2023,onuchicramos2025} and to the shifting of responsibility under delegation \citep{bartling2012blame,chassang2024secure}.

Consider completed projects with a predetermined team of \(n\geq 1\) authors that used AI with a fixed probability across team sizes. Conditional on use, one author operated the tool with prior probability \(1/n\), and silence leaves readers with posterior probability \(r_n\) that a silent paper used AI. Every author bears an undivided personal cost \(b\geq 0\) from withholding a truthful statement, which can stand for honesty norms or an outlet's expectations, and a reputational loss of \(s>0\) times the reader's posterior probability that this author operated the tool. An unattributed disclosure assigns operator probability \(1/n\) and silence assigns \(r_n/n\), so each author's net payoff from disclosing rather than staying silent is \begin{equation}
 \Delta u_n=b-\frac{s}{n}(1-r_n).
 \label{eq:attribution-payoff}
\end{equation} Appendix C.3 derives the unique rational-expectations cutoff equilibrium and shows that the disclosure threshold on \(b\) falls strictly as \(n\) rises. The personal cost of silence stays at \(b\) for every collaborator, whereas the incremental stigma \((s/n)(1-r_n)\) shrinks with team size.

Under homogeneous preferences, the model predicts that disclosing papers average more authors than silent papers even when AI leaves every roster unchanged. Across assessable post-2022 papers, disclosing papers average only marginally more authors than silent ones, and both medians equal three. For the main estimates the important point is what the model does not predict. It supplies no direct prediction for research assistants or page length. Attribution selection could generate assistant or page differences only if larger author teams already used more assistants and wrote longer papers while the rest of the staffing was unrelated to reporting. Heterogeneous attitudes and a unanimity rule can even reverse the author gradient, because a larger team contains more possible objectors.

\begin{figure}[htbp]\centering
\includegraphics[width=\linewidth]{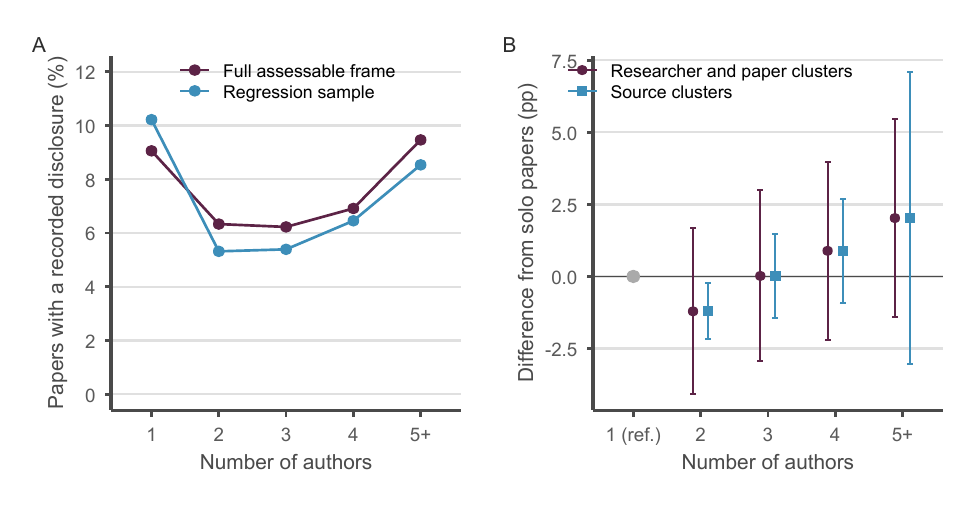}
\caption{Recorded disclosure by number of authors}\label{fig:teamsize}
\tabnote{Panel A: share of timing-supported post-2022 papers with a recorded own-use disclosure, by author count, in the full assessable frame (16,300 papers, 1,167 disclosures) and in the author-count regression sample (8,502 papers, 544 disclosures). Panel B: within-researcher differences in disclosure probability relative to solo papers, in percentage points, from a linear probability model with researcher, source-quarter, and field effects, the writing flag, and prose controls, on the 8,502-paper sample; 95\% intervals clustered by researcher and paper (circles) and by source (squares). Disclosure information is concentrated in about four effective sources. No recorded statement does not establish non-use.}
\end{figure}

Figure~\ref{fig:teamsize} compares raw disclosure rates across team sizes with within-researcher differences relative to solo papers. In the raw data, solo papers disclose AI use more often than papers with any larger team. That solo excess contradicts the homogeneous model's pooled prediction. But it disappears once researcher fixed effects enter. Within the same economist's portfolio, no team size differs significantly from solo papers in its disclosure probability, and a doubling of the author count is associated with no detectable change.

A mechanical benchmark offers a separate check on how disclosures aggregate across coauthors. If each author independently triggered a disclosure with the same probability within a source-year cell, a complementary log-log regression, a link function under which independent union probabilities are linear in log team size, would give a slope of one on \(\log n\). The estimated slope is far below one. The data reject identical independent author-level reporting without ruling out heterogeneous task roles. Before estimation, I also froze eight competing mechanism tests, reported in Appendix B.6: coauthor field overlap and citation standing, coauthors' pre-2022 documentation habits, outside-team propensities, detection-recall bounds, submission-preparation signals, and protective wording (Appendix B.5). Coauthored disclosures use protective wording somewhat more often, but no test survives the Holm adjustment. The estimable tests rest on very few effective sources and separate neither attribution nor documentation habits from actual AI use.

\section{AI indicators, research assistants, and project scope}\label{sec:results}

\subsection{Main estimates}\label{main-estimates}

Table 1 reports within-researcher estimates from Equation 2 for the two AI indicators across four primary outcomes: any credited research assistant, the count of named assistants, the count of byline authors, and total PDF pages. Both indicators enter together with the prose controls and the researcher, source-quarter, and field fixed effects. Papers without a disclosure in the estimation sample average 0.83 named assistants and 52.4 pages. Reading across the two columns asks whether an explicit AI statement and AI-like prose move with human inputs in the same way.

\begin{table}[htbp]\centering\caption{Within-researcher associations between AI indicators and paper inputs}\label{tab:main}
{\small\renewcommand{\arraystretch}{1.12}\setlength{\tabcolsep}{4pt}\begin{tabular*}{\linewidth}{@{\extracolsep{\fill}}lrrrrr@{}}\toprule
Outcome & Writing flag & Disclosed & Papers & \shortstack{Disclosed\\papers} & Researchers\\\midrule
Any assistant (pp) & \begin{tabular}[t]{@{}c@{}}-3.206\\(2.251)\end{tabular} & \begin{tabular}[t]{@{}c@{}}6.599\\(2.606)\end{tabular} & 7,243 & 468 & 4,604\\
Named assistants & \begin{tabular}[t]{@{}c@{}}-0.131\\(0.111)\end{tabular} & \begin{tabular}[t]{@{}c@{}}0.548\\(0.168)\end{tabular} & 7,242 & 468 & 4,603\\
Authors & \begin{tabular}[t]{@{}c@{}}0.011\\(0.061)\end{tabular} & \begin{tabular}[t]{@{}c@{}}0.240\\(0.093)\end{tabular} & 8,502 & 544 & 5,285\\
Pages & \begin{tabular}[t]{@{}c@{}}-0.693\\(1.100)\end{tabular} & \begin{tabular}[t]{@{}c@{}}5.346\\(1.274)\end{tabular} & 8,502 & 544 & 5,285\\
\bottomrule\end{tabular*}}\tabnote{Estimates are from Equation 2 with the writing flag and disclosure entered together, prose-length and window controls, researcher, source-quarter, and field fixed effects, and paper-total weights. Standard errors clustered by researcher and paper in parentheses; any-assistant estimates are in percentage points. Across the twenty-test family (two indicators by ten outcomes), the disclosure p-values are: any assistant raw p = .011, Romano--Wolf p = .157, Holm p = .170; named assistants raw p = .001, Romano--Wolf p = .020, Holm p = .019; authors raw p = .010, Romano--Wolf p = .147, Holm p = .163; pages raw p \textless{} .001, Romano--Wolf p = .0005, Holm p = .0006. All ten writing-flag tests have Romano--Wolf p above .64. Neither adjustment accounts for the 3.3 effective sources.}\end{table}

Within an economist's portfolio, papers that disclose AI use credit more named research assistants and run longer. Disclosing papers credit 0.548 additional named assistants, two-thirds of the mean on papers without a disclosure, and the association survives the twenty-test adjustment under both Romano--Wolf and Holm. They also run 5.35 pages longer, about a tenth more than the baseline, an even more precisely estimated difference. Disclosing papers are also more likely to credit any assistant and carry slightly more authors. Both differences are individually significant, but neither survives the family-wise adjustment, and neither carries weight in what follows.

Resemblance to AI-generated writing shows a different pattern in the first column. Flagged papers differ little from unflagged papers by the same economist on any of the four outcomes, and every interval includes zero. In pooled regressions without researcher intercepts (Appendix Table~\ref{tab:full}), flagged papers credit noticeably less assistance; comparing projects within the same economist shrinks that gap towards zero. Read through Equation 1, the two columns fit a technology that expands scope when applied to research operations and leaves staffing untouched when confined to separable prose editing.

\subsection{Which disclosures carry the association}\label{which-disclosures-carry-the-association}

If scope expansion explains the positive assistant estimate in Table 1, disclosures that report research operations should carry the association and statements confined to prose editing should look like the writing flag. I test that implication in an exploratory split outside the registered twenty-test family and report raw p-values. Using the whole-document task rule from Section 3, disclosures divide into operational or mixed statements, writing-only statements, and unlabelled statements, the last controlled separately. Operational disclosures outnumber writing-only ones by about five to one in every sample, so the writing-only estimates are less precise.

Table 2 reports the within-researcher coefficients, standard errors, raw p-values, and the test of equality between the operational and writing-only coefficients for the four main outcomes.

\begin{table}[htbp]\centering\caption{Exploratory within-researcher estimates by the disclosed AI task mix}\label{tab:taskmix}
{\small\renewcommand{\arraystretch}{1.12}\setlength{\tabcolsep}{4pt}\begin{tabular*}{\linewidth}{@{\extracolsep{\fill}}lrrr@{}}\toprule
Outcome & Operational & Writing-only & Equality $p$\\\midrule
Any assistant (pp) & \begin{tabular}[t]{@{}c@{}}6.9 (2.9)\\.020\end{tabular} & \begin{tabular}[t]{@{}c@{}}1.7 (6.0)\\.78\end{tabular} & .43\\
Named assistants & \begin{tabular}[t]{@{}c@{}}0.679 (0.199)\\.0007\end{tabular} & \begin{tabular}[t]{@{}c@{}}-0.098 (0.209)\\.64\end{tabular} & .0034\\
Authors & \begin{tabular}[t]{@{}c@{}}0.300 (0.103)\\.0035\end{tabular} & \begin{tabular}[t]{@{}c@{}}0.056 (0.235)\\.81\end{tabular} & .32\\
Pages & \begin{tabular}[t]{@{}c@{}}6.78 (1.44)\\$<.0001$\end{tabular} & \begin{tabular}[t]{@{}c@{}}0.83 (2.97)\\.78\end{tabular} & .071\\
\bottomrule\end{tabular*}}\tabnote{Entries are the coefficient, the researcher/paper-clustered standard error in parentheses, and the raw p-value from replacing the single disclosure indicator in Equation 2 with operational-or-mixed, writing-only, and unlabelled indicators; the final column reports the raw p-value of a test that the operational and writing-only coefficients are equal. Samples match Table 1, with 361 operational, 68 writing-only, and 39 unlabelled disclosures for assistance and 425, 78, and 41 for authors and pages. All tests lie outside the registered twenty-test family.}\end{table}

Operational disclosures account for the whole named-assistant association. Within a researcher's portfolio, papers that report coding, data, analytical, or mathematical AI use credit 0.679 more named assistants, whereas papers that report only drafting, editing, translation, or typesetting credit no more assistants than the same researcher's silent papers. The two coefficients differ clearly. The contrast matters for the reporting explanation. Writing-only authors also chose to place an AI statement in their paper, so a general willingness to document every input would raise assistant counts in both columns, not only in the operational one.

Document length follows the same division. Operational disclosures run several pages longer within researcher, whereas writing-only disclosures are barely longer than silent papers, although the test of equality is only marginally significant. The raw means tell the same story: operational papers credit the most assistants and run longest, and writing-only papers sit slightly below papers without a disclosure on both. Pooled regressions without researcher intercepts also separate the two types on named assistants and pages, but not on authors or incidence. The writing-only cells are modest, and only a handful of economists have papers in both task categories, so the comparison rests mainly on contrasting each disclosure group with the silent papers of the same scholar.

\subsection{Where the extra pages are}\label{where-the-extra-pages-are}

Five extra pages on a disclosing paper could be padded main text or additional supporting analyses, robustness tables, and data documentation in an appendix. To locate the added length I run a second exploratory analysis outside the registered family, with raw p-values, on almost all of the pages-sample manuscripts, those with verified page text. The boundary rule takes the first appendix page from usable PDF bookmarks where they exist and otherwise from a standalone appendix heading or a lettered section that follows the last numbered section or the bibliography. On a held-out validation sample, the rule places the appendix boundary correctly in about nine papers in ten; Appendix B reports the error rates.

Table 3 splits total length into main-text pages, appendix pages, body pages before both references and appendix, and an indicator for having any appendix.

\begin{table}[htbp]\centering\caption{Exploratory within-researcher decomposition of paper length into main text and appendix}\label{tab:pagesplit}
{\small\renewcommand{\arraystretch}{1.12}\setlength{\tabcolsep}{4pt}\begin{tabular*}{\linewidth}{@{\extracolsep{\fill}}>{\raggedright\arraybackslash}p{2.1in}rrr@{}}\toprule
Outcome & Disclosed & Operational & Writing-only\\\midrule
Total pages & \begin{tabular}[t]{@{}c@{}}5.29 (1.28)\\$<.001$\end{tabular} & \begin{tabular}[t]{@{}c@{}}6.70 (1.45)\\$<.0001$\end{tabular} & \begin{tabular}[t]{@{}c@{}}0.81 (2.97)\\.79\end{tabular}\\
Main-text pages & \begin{tabular}[t]{@{}c@{}}1.63 (0.70)\\.020\end{tabular} & \begin{tabular}[t]{@{}c@{}}1.70 (0.76)\\.026\end{tabular} & \begin{tabular}[t]{@{}c@{}}3.02 (1.83)\\.098\end{tabular}\\
Appendix pages & \begin{tabular}[t]{@{}c@{}}3.66 (1.10)\\.0009\end{tabular} & \begin{tabular}[t]{@{}c@{}}5.00 (1.28)\\.0001\end{tabular} & \begin{tabular}[t]{@{}c@{}}-2.22 (2.27)\\.33\end{tabular}\\
Body pages (before references and appendix) & \begin{tabular}[t]{@{}c@{}}1.16 (0.61)\\.059\end{tabular} & \begin{tabular}[t]{@{}c@{}}1.07 (0.67)\\.11\end{tabular} & \begin{tabular}[t]{@{}c@{}}2.63 (1.65)\\.11\end{tabular}\\
Any appendix (pp) & \begin{tabular}[t]{@{}c@{}}3.4 (2.2)\\.11\end{tabular} & \begin{tabular}[t]{@{}c@{}}2.3 (2.1)\\.26\end{tabular} & \begin{tabular}[t]{@{}c@{}}1.9 (4.5)\\.67\end{tabular}\\
\bottomrule\end{tabular*}}\tabnote{Estimates use the 8,493-paper verified-text subset with 542 disclosures (the full-sample total-page estimate in Table 1 is 5.35), the Equation 2 controls and fixed effects, researcher/paper-clustered standard errors in parentheses, and raw p-values. Column 1 uses the single disclosure indicator; columns 2 and 3 come from the task-mix specification. All estimates lie outside the registered twenty-test family.}\end{table}

About seven-tenths of the extra length on disclosing papers sits in the appendix and three-tenths in the main text. Disclosing papers are not much more likely to have an appendix at all; their appendices are longer. Operational disclosures lean even more on the back matter, whereas writing-only disclosures show no appendix gain. The appendix share of the extra length holds when NBER or arXiv is dropped and among papers whose boundaries are confirmed by bookmarks. The writing flag, by contrast, points towards shorter appendices and slightly longer main text, a pattern that is suggestive at most.

\subsection{Documentation and source composition}\label{documentation-and-source-composition}

\begin{figure}[htbp]\centering
\includegraphics[width=\linewidth]{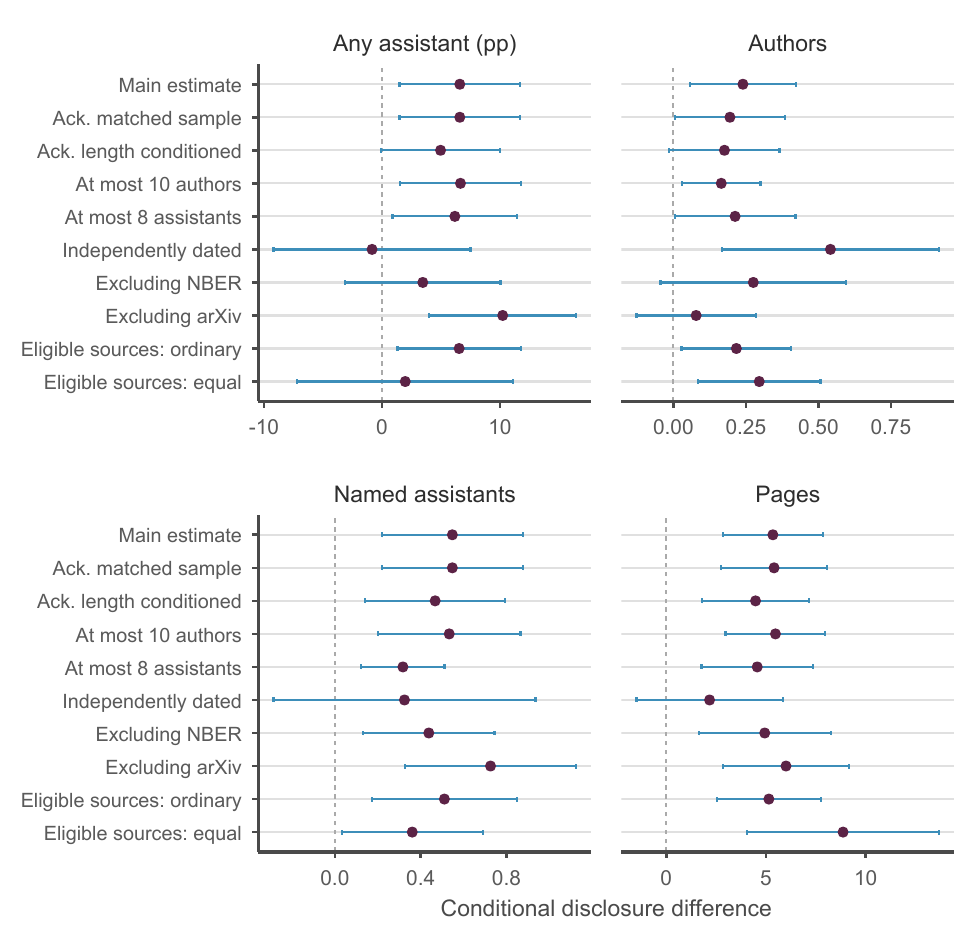}
\caption{Disclosure, documentation, and source composition}\label{fig:sensitivity}
\tabnote{Disclosure coefficients with pointwise 95\% researcher/paper-clustered intervals, conditional on that dependence structure. Acknowledgement-length conditioning is compared with its same-sample baseline. Equal source weights are compared with ordinary weights on the same eligible-source subset. Both AI indicators and the fixed effects in equation~\ref{eq:empirical} remain included. Figure~\ref{fig:writing} uses the same specification rows for three writing indicators and two team outcomes. Source concentration and overlapping authors limit interpretation of the intervals; see Appendix~\ref{sec:inference}.}
\end{figure}

Figure~\ref{fig:sensitivity} places the main disclosure coefficients beside alternative functional forms, controls for acknowledgement space, and source-weighting schemes. Conditioning on the acknowledgement word count attenuates both assistant estimates modestly on identical samples. Thanking assistants and stating AI use both add words to an acknowledgement, so conditioning on its length absorbs part of the outcome as well as general verbosity. It is a descriptive check, not a bias correction.

Large or skewed assistant rosters contribute part, though not all, of the headcount magnitude. The log of named assistants and a count capped at eight names both remain strongly significant across the twenty-test family. Dropping papers that credit more than eight assistants shrinks the linear estimate by about two-fifths (Appendix B.2), so ordinary teams and large lab-style rosters both contribute. A separate count of documented empirical and theoretical methods fails the adjustment (Appendix B.4).

Outlet composition shapes binary assistant incidence far more than named headcounts or page length. Omitting NBER makes the incidence coefficient smaller and imprecise, whereas omitting arXiv, where acknowledgements rarely list junior staff, makes it larger. Weighting outlets equally rather than by paper reduces incidence to an imprecise estimate near zero and trims the named-assistant estimate. Dropping any single source leaves the page difference essentially unchanged, while the author difference shrinks when arXiv is dropped. Restricting links to work-corroborated researcher identities keeps all four disclosure coefficients positive with a smaller author estimate. The small independently dated subset gives larger author and smaller page differences, and no incidence difference.

\subsection{Writing resemblance}\label{sec:writing-comparison}

\begin{figure}[htbp]\centering
\includegraphics[width=\linewidth]{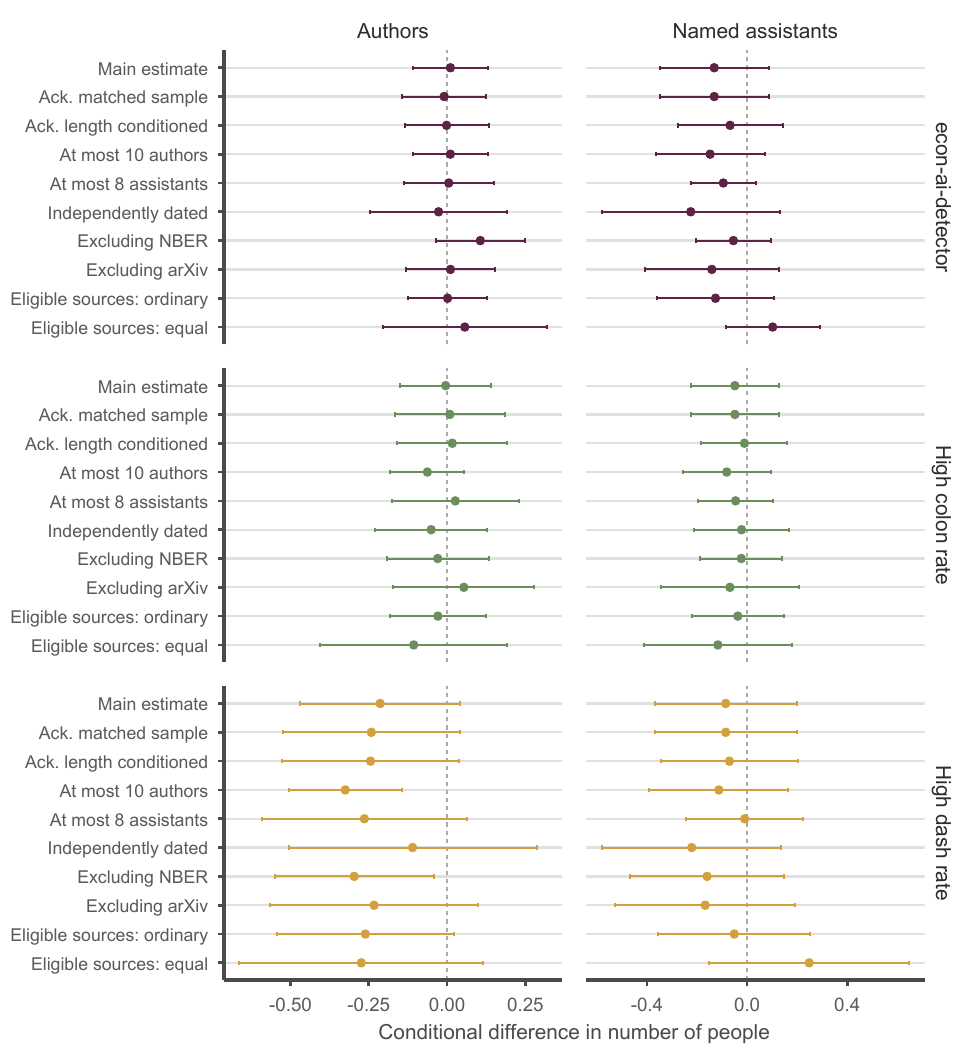}
\caption{Writing indicators, authors, and named assistants}\label{fig:writing}
\tabnote{Coefficients with pointwise 95\% researcher/paper-clustered intervals. Each writing indicator enters separately with disclosure and the controls and fixed effects in equation~\ref{eq:empirical}. Corresponding rows use identical samples and weights across indicators. High colon and dash rates exceed the historical pooled mean plus two standard deviations (8.905 and 9.267 per 1,000 words). Equal-source comparisons retain sources with at least five disclosed and twenty no-statement papers. Scales match within each outcome. These are writing proxies, not verified AI use; specifications overlap and uncertainty remains conditional on the stated dependence structure.}
\end{figure}

Figure~\ref{fig:writing} plots within-researcher estimates with pointwise 95\% intervals for author and named-assistant counts across ten sensitivity specifications and three text flags: the strict \texttt{econ-\allowbreak ai-\allowbreak detector} indicator, a high colon rate, and a high dash rate. Every specification pairs one text indicator with the disclosure and the Equation 2 controls on identical samples. Across the ten specifications, neither the detector flag nor the colon rate shows a consistent association with author or named-assistant counts: every detector and colon interval includes zero. The dash rate gives a negative author association in two specifications, under the ten-author cap and when NBER is omitted; all other dash intervals span zero.

The intervals in the main detector specification are nevertheless wide enough to permit moderate reductions in staffing, especially in named assistants. The detector estimates therefore do not rule out team contractions of the size \citet{wang2026} report in other disciplines with different prose classifiers. Measurement error in automated detection \citep{jabarianimas2025} and unrecorded AI use among silent papers can both attenuate the prose-flag comparisons. But the contrast with the operational disclosures matches the separable-writing case of Equation 1. The prose a detector reads is a measure of how the paper was written, not of the argument or evidence beneath it \citep{fourie2026writes}.

Appendix A reports supplementary reference-list and university-affiliation comparisons. Disclosing papers cite more recent references, but the gap more than halves once papers about AI itself are excluded, and they cite no more references in total. Conditioning on university-tier shares leaves the any-assistant and page estimates essentially unchanged.

\section{Apprenticeship, portfolios, and the division of research labour}\label{apprenticeship-portfolios-and-the-division-of-research-labour}

Within a recurring economist's portfolio, projects that disclose operational AI credit more named research assistants and run longer than silent papers by the same scholar. The institutional series show a matching aggregate pattern: the share crediting any assistant holds steady in every half-year from 2022 to 2026 as disclosure rises. On the finished manuscript, the junior rung of academic economics has not disappeared with the arrival of language models. But paper-level regressions speak only to acknowledged contributions on completed working papers by recurring scholars. They do not observe department-wide hiring, junior wages, contract durations, or projects abandoned before circulation, and they exclude economists who post too rarely to enter the sample.

The distinction between comparing across economists and comparing within one economist's portfolio reconciles these findings with recent evidence on unequal gains from AI. \citet{jiang2026equalizing} find across economists that generative AI narrowed the productivity gap between researchers with and without research assistants, raising the top-journal publication share of unassisted authors. Substitution across persons and expansion within a portfolio are distinct margins of the same technology. For a scholar working without human support, a language model performs the routine coding and data cleaning that once required grant money to hire an assistant. Within an assisted economist's portfolio, cheaper execution lowers the operational cost in Equation 1 and makes wider studies, with fuller appendices and more credited names, worth undertaking \citep{acemoglu2019}.

Even inside an assisted research group, more credited names per paper need not mean more junior jobs or more paid hours. An acknowledgement records a person's contribution to a paper, not an employment contract or a time sheet. One full-time pre-doctoral fellow can spend a year on a single paper and next year contribute shorter tasks to three papers, raising credits per paper at unchanged headcount. A wider investigation can also add a part-time student for one specialised data check while total assistant hours in the group fall. Paper-level estimates therefore leave open how an economist's total demand for assistants evolves across an annual portfolio.

Appendix B.7 aggregates manuscripts into complete annual researcher portfolios across the series present throughout the window, without conditioning on disclosure. Mean credited assistants per paper dipped in 2023 and then rose above its 2022 level by 2025. Whether that increase reflects new hiring or wider reuse across papers turns on how many distinct people stand behind a scholar's annual credits. Acknowledgements print names without affiliations or dates, so matching strings across two papers can mean one reused assistant or two people who share a name. Identity-free bounds, set between a scholar's largest single-paper roster in that year and the total annual credits, place distinct assistants per portfolio only modestly above one in 2025. Coauthors' portfolios also overlap on joint manuscripts, so researcher-level counts cannot be summed into a discipline-wide workforce.

Task allocation inside those portfolios matters beyond short-run headcounts because research assistance is a central apprenticeship of economics. Novices clean raw datasets and write estimation code in exchange for supervision and for letters that carry weight in doctoral admissions. \citet{garicanorayo2025} show that on-the-job training survives only above a viability threshold: a junior's productive contribution on delegated tasks must cover the senior's cost of supervising and teaching. Once AI absorbs enough routine work that doing the task oneself costs a senior less than instructing a novice, the surplus that finances mentorship disappears and the apprenticeship breaks down. \citet{wang2026predoc} models the pre-doctoral labour market directly and shows that AI changes both the execution value of pre-doctoral fellows and the signal their work sends in admissions. \citet{emanuel2026} find that proximity to experienced coworkers raises the feedback junior workers receive.

Steady institutional assistant shares and the operational disclosure association are consistent with a junior position that persists on paper. Credits cannot reveal where that position sits relative to the viability threshold. Identical thank-you lines can cover different bundles of junior work as job boundaries shift \citep{autor2025,garicano2026bundles,gans2026bundling}. On one disclosing project an assistant may link archival sources and run supplementary estimations that cheaper computation made worth pursuing. On another the assistant may check model-generated scripts and audit appendix numbers produced from a senior author's prompts \citep{agrawal2026}. Building an empirical pipeline from raw files teaches a novice to spot data anomalies and identification threats; reviewing fluent machine output without writing the code may teach verification without that foundation. Corroborated assistant identities and hours across portfolios would separate hiring from reuse. Project-level records of task assignments and faculty feedback would separate routine execution from verification.

\section{Conclusion}\label{sec:conclusion}

Within a recurring economist's portfolio, working papers that disclose generative AI credit more named research assistants and run longer than silent papers by the same scholar. Statements that report coding, data preparation, analysis, or derivation carry both associations, and most of the added length sits in appendices rather than main text. Writing-only disclosures and AI-like prose show no consistent association with authors, named assistants, or pages, and the positive point estimates for authors and for the probability of any assistant fail the family-wise adjustment. Among the projects that recurring economists circulate as working papers, operational AI currently accompanies wider, assistant-supported investigations rather than assistantless ones.

Bringing acknowledgements into the study of scientific teams connects the debate over the displacement of junior knowledge workers to the one research setting that records junior labour next to finished output. Payroll and job-posting studies find slower entry-level hiring in corporate occupations where AI automates routine work. The institutional economics series pair rising AI disclosure with a steady share of papers crediting assistants and with larger rosters on operational projects. The weak overlap between the two AI indicators adds a lesson for measuring AI in science: voluntary disclosure and prose resemblance are different objects. A prose classifier detects drafting style where staffing stays flat; an explicit disclosure records an operational workflow, filtered through a team's reporting incentives.

The findings describe credited contributions to completed manuscripts in 24 selected working-paper series, not paid hours, wages, distinct headcounts, or department-wide employment, and the disclosure variation rests on about 3.3 effective sources. A disclosure is a collective decision by an author team rather than an instrument reading of actual use. Researcher fixed effects and the contrast between operational and writing-only statements narrow the simple documentation explanations, but unobserved project ambition and selective silence among non-disclosing papers remain part of every comparison.

Two designs can separate the mechanisms that working-paper archives bundle together. An experiment that holds an AI-assisted task fixed and varies whether the disclosure names the operator or speaks for the whole team would test whether coauthorship dilutes personal stigma. Inside active research groups, linking project logs to assistants' time and task records would separate reuse across papers from new hiring. Before a working paper reaches a series, a language model can now clean the raw files, write the estimation code, search the literature, and assemble the appendix tables in minutes. When AI takes that first pass, does the research assistant thanked behind the byline build broader evidence or check machine output? And, more importantly, does checking still teach a novice how to do economics?

\FloatBarrier\clearpage\begingroup\small\setstretch{1.1}\setlength{\bibsep}{3pt}\bibliographystyle{v16/aea}\bibliography{v16/references}\endgroup
\FloatBarrier\clearpage\appendix
\renewcommand{\thetable}{\thesection\arabic{table}}
\renewcommand{\thefigure}{\thesection\arabic{figure}}
\setcounter{table}{0}\setcounter{figure}{0}
\section{Data, measures, and samples}\label{app:data}
\subsection{Source coverage and document timing}
Institutional screening considered established RePEc-listed outlets, geographical coverage, and annual catalogue volume. Annual catalogue volume, including a ten-item benchmark, was a screening preference rather than a uniform eligibility threshold. Selected-family and regression counts can be smaller. Penn PIER is excluded because its retained post-2022 coverage consists of one paper. The accessible Swiss Finance Institute copies do not provide coverage on both sides of 2022. The numbered Bank of England Staff Working Papers and Boston Fed Research Department Working Papers are reconciled within the sample window. Publisher withdrawals and unavailable source files retain explicit missing-source dispositions. Catalogue enumeration does not establish that every paper has retrievable, date-supported text. The Central Bank of Chile, Pretoria, Chicago BFI, Paris School of Economics, Tokyo CIRJE, Toronto, Yale Cowles, and Stanford SIEPR series could not be completely acquired from public sources and are excluded to avoid selective coverage. The coverage rule requires at least 98\% acquisition overall and in each year from 2023 to 2026. Bank of England and Philadelphia Fed are retained exceptions: their 2026 coverage is 47/48 and 46/47 papers, respectively, each one paper short of complete coverage. The retained series and family counts appear in Table~\ref{tab:coverage}. The sources are a selected set of outlets, not a representative sample of economists.

Table~\ref{tab:coverage} separates retained records, supported timing, and observed measures. Acquisition and deduplication mean that retained counts can differ from catalogue counts. In particular, an absent retained family in a source-year is not evidence that the outlet issued no papers. Unavailable PDFs and unresolved extraction remain missing in the affected measure. Exact hashes and corroborated title/author information identify candidate versions and cross-posts; date and source priorities choose representatives without replacing a record because of a missing outcome. Residual cross-posts can remain, although no fitted outcome contains two records with an identical PDF.

\begin{figure}[htbp]\centering
\includegraphics[width=\linewidth]{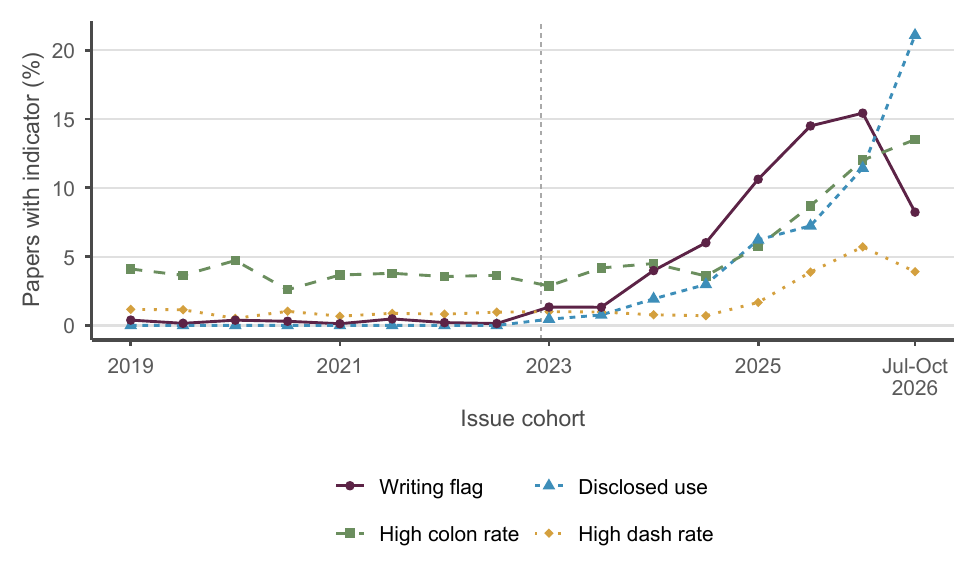}
\caption{Trends in AI indicators}\label{fig:visibility}
\tabnote{Half-year positive-indicator shares among timing-supported papers, using each measure's assessable denominator. The last observations cover July--October 2026, an incomplete half-year. High punctuation denotes rates above fixed historical thresholds. Issue dates need not date every word in a retrieved revision; punctuation includes some non-prose marks. Definitions appear in Appendix~\ref{app:data}.}
\end{figure}

The frame comprises 11,160 independently version-dated documents, 31,303 additional documents included under the printed-date assumption, and 14,122 with unresolved or later-revision timing. The first two groups constitute the 42,463 timing-supported records. An independently dated copy supplies evidence about when the measured text existed. A printed date consistent with its issue cohort instead requires no unrecorded later revision affecting the measures; a file hash alone cannot establish this. Explicit later revisions, conflicting dates, and undated copies do not automatically supply historical outcomes. Repository citation sheets do not substitute for dates on manuscript front matter. A hypothetical publisher notice that a paper may be revised is not evidence of an actual revision. A bare manuscript-version date remains a printed-date observation; it does not establish an independently verified first edition. Actual revision notices and later front-matter dates retain the same timing restrictions. Figure~\ref{fig:visibility} ends with July--October 2026, an incomplete half-year.

\begingroup\small\setlength{\tabcolsep}{4pt}\setlength{\LTleft}{0pt}
\begin{longtable}{lrrrrr}
\caption{Coverage by retained source}\label{tab:coverage}\\
\toprule
Series & Records & Dated & Scored & Assistance & Disclosed\\\midrule\endfirsthead
\multicolumn{6}{@{}l}{\small Table~\ref{tab:coverage} continued}\\
\toprule
Series & Records & Dated & Scored & Assistance & Disclosed\\\midrule\endhead
arXiv economics & 17,204 & 7,332 & 12,824 & 11,288 & 1,010\\
Bank of Canada & 641 & 599 & 617 & 628 & 4\\
BIS & 823 & 699 & 741 & 771 & 19\\
Bank of England & 656 & 541 & 594 & 606 & 3\\
Warwick CAGE & 588 & 514 & 491 & 553 & 15\\
Cambridge & 929 & 772 & 786 & 770 & 14\\
US Census CES & 654 & 355 & 485 & 577 & 3\\
LSE CEP & 709 & 611 & 648 & 702 & 17\\
CESifo & 7,284 & 6,377 & 6,586 & 6,066 & 172\\
Osaka ISER & 421 & 366 & 367 & 385 & 11\\
ANU CAMA & 840 & 798 & 763 & 667 & 8\\
Boston Fed & 212 & 139 & 181 & 195 & 2\\
San Francisco Fed & 282 & 172 & 232 & 252 & 4\\
Federal Reserve Board & 1,072 & 807 & 908 & 943 & 14\\
Chicago Fed & 249 & 153 & 206 & 214 & 9\\
Institute for Fiscal Studies & 358 & 249 & 326 & 341 & 10\\
NBER & 14,836 & 13,876 & 11,723 & 13,919 & 345\\
New York Fed & 477 & 295 & 410 & 432 & 5\\
Philadelphia Fed & 425 & 378 & 355 & 376 & 4\\
RIETI & 1,256 & 1,144 & 1,058 & 1,246 & 8\\
Tinbergen & 968 & 872 & 894 & 780 & 10\\
Toulouse & 974 & 846 & 817 & 840 & 7\\
Pompeu Fabra & 414 & 279 & 340 & 371 & 4\\
World Bank & 4,313 & 4,289 & 3,479 & 3,750 & 31\\
\midrule
Total & 56,585 & 42,463 & 45,831 & 46,672 & 1,729\\
\bottomrule\end{longtable}

\noindent Records are selected paper-family representatives. Dated denotes timing-supported documents. Scored and assistance count assessable writing and assistant-incidence observations in the full source frame; they need not be timing-supported. Disclosed counts positive statements after the second review described in Section~\ref{sec:measurement-quality}, not documents searched. These columns therefore have different denominators and are not successive stages of a single sample restriction.\par\endgroup

\begingroup\small\setlength{\parindent}{0pt}\setlength{\tabcolsep}{4pt}\setlength{\LTleft}{0pt}\begin{longtable}{lrrrrr}
\caption{Half-year disclosure and assistant credits}\label{tab:adoption}\\\toprule
Group & Half-year & Disclosed (\%) & $N_D$ & Assistant (\%) & $N_A$\\\midrule
\endfirsthead\toprule
Group & Half-year & Disclosed (\%) & $N_D$ & Assistant (\%) & $N_A$\\\midrule
\endhead
All sources & 2022 H1 & 0.00 & 1,734 & 28.05 & 1,565\\
arXiv & 2022 H1 & 0.00 & 314 & 9.09 & 209\\
Institutional & 2022 H1 & 0.00 & 1,420 & 30.97 & 1,356\\
NBER & 2022 H1 & 0.00 & 517 & 41.33 & 525\\
All sources & 2022 H2 & 0.00 & 1,712 & 27.52 & 1,530\\
arXiv & 2022 H2 & 0.00 & 281 & 7.57 & 185\\
Institutional & 2022 H2 & 0.00 & 1,431 & 30.26 & 1,345\\
NBER & 2022 H2 & 0.00 & 551 & 38.61 & 562\\
All sources & 2023 H1 & 0.45 & 1,766 & 24.71 & 1,574\\
arXiv & 2023 H1 & 1.25 & 320 & 6.67 & 195\\
Institutional & 2023 H1 & 0.28 & 1,446 & 27.27 & 1,379\\
NBER & 2023 H1 & 0.55 & 547 & 32.26 & 558\\
All sources & 2023 H2 & 0.76 & 1,710 & 26.08 & 1,507\\
arXiv & 2023 H2 & 1.18 & 340 & 5.91 & 220\\
Institutional & 2023 H2 & 0.66 & 1,370 & 29.53 & 1,287\\
NBER & 2023 H2 & 0.92 & 543 & 37.48 & 555\\
All sources & 2024 H1 & 1.93 & 1,812 & 25.76 & 1,615\\
arXiv & 2024 H1 & 2.17 & 368 & 5.69 & 246\\
Institutional & 2024 H1 & 1.87 & 1,444 & 29.36 & 1,369\\
NBER & 2024 H1 & 2.61 & 574 & 36.77 & 582\\
All sources & 2024 H2 & 2.98 & 1,949 & 26.66 & 1,714\\
arXiv & 2024 H2 & 3.58 & 419 & 7.50 & 280\\
Institutional & 2024 H2 & 2.81 & 1,530 & 30.40 & 1,434\\
NBER & 2024 H2 & 3.76 & 611 & 36.92 & 623\\
All sources & 2025 H1 & 6.21 & 1,963 & 26.78 & 1,703\\
arXiv & 2025 H1 & 7.86 & 509 & 6.94 & 317\\
Institutional & 2025 H1 & 5.64 & 1,454 & 31.31 & 1,386\\
NBER & 2025 H1 & 8.18 & 599 & 40.49 & 610\\
All sources & 2025 H2 & 7.24 & 2,211 & 24.01 & 1,841\\
arXiv & 2025 H2 & 6.99 & 730 & 7.54 & 451\\
Institutional & 2025 H2 & 7.36 & 1,481 & 29.35 & 1,390\\
NBER & 2025 H2 & 8.15 & 577 & 39.35 & 582\\
All sources & 2026 H1 & 11.43 & 2,694 & 22.00 & 2,164\\
arXiv & 2026 H1 & 13.21 & 1,037 & 6.53 & 597\\
Institutional & 2026 H1 & 10.32 & 1,657 & 27.89 & 1,567\\
NBER & 2026 H1 & 13.84 & 701 & 32.46 & 724\\
All sources & 2026 H2* & 21.08 & 2,196 & 21.01 & 1,666\\
arXiv & 2026 H2* & 24.58 & 1,139 & 5.58 & 699\\
Institutional & 2026 H2* & 17.31 & 1,057 & 32.16 & 967\\
NBER & 2026 H2* & 18.09 & 503 & 40.57 & 493\\
\bottomrule\end{longtable}\tabnote{Timing-supported papers, with outcome-specific assessable denominators. Shares are percentages; counts are denominators. Institutional includes NBER. *July--October 2026 is incomplete.}\par\endgroup

\FloatBarrier

\subsection{Measurement and researcher identities}
Assistance incidence follows a literal research-assistance role rule, with named counts requiring evidence identifying both person and role. Readers and advisers are recorded separately. Broader data, execution, and coordination support also enters the definition of solo authorship without assistance. Ambiguous role phrases remain uncertain rather than becoming negative observations. A collective credit can establish assistance while no individuals are named: zero named people does not imply zero total people. Incomplete named-count evidence remains missing. Negative incidence requires usable acknowledgement coverage; empty or unreadable extraction does not supply a zero.

The writing implementation pins \texttt{econ-ai-detector} commit \texttt{11285447a3cc} and version-3 neural revision \texttt{fb7fb8b27bb9}. A paper is flagged when its second-highest logistic window score is at least 0.33489 and its second-highest neural logit margin is at least 7.4827. The other released threshold pairs are $(0.18983,-3.7104)$ and $(0.14309,-5.4591)$. These are calibration choices, not guaranteed error rates in economics working papers. The continuous score is the second-highest logistic score, expressed in changes of 0.1 where used. Unsupported language, insufficient prose, and unresolved section selection remain missing.

Own-use disclosures require statements about generative AI's contribution to producing the paper. Full-text tool and model terms identify candidate passages for contextual classification, which distinguishes own use, AI as a research topic, conventional tools, explicit non-use, and ambiguity. Positive evidence is checked against a literal source passage. Translation, writing, coding, search, data preparation, analysis, derivation, and checking labels can overlap. Under the whole-document task rule, a disclosure is writing-only when every supported task concerns manuscript drafting, editing, translation, or typesetting; research coding, data annotation, substantive search, estimation, derivation, or the generation of experimental inputs make it operational or mixed, and unspecified brainstorming or feedback is unclear rather than writing-only. Models that are studied, benchmarked, or supplied to experimental participants do not establish own use. Generic AI, or a grammar or translation service without evidence of generative functionality, remains unresolved. Saved prompts and per-call model identifiers accompany the replication code; Section~\ref{sec:measurement-quality} describes the classification procedures and their model-provenance limits. This is measurement from text with explicit coding rules, not an inference of actual use from missing statements \citep{ludwig2025}.

Each confirmed disclosure passage was also coded for its content. Protective wording denotes an explicit restriction on AI's intellectual role or a claim that humans reviewed, checked, validated, or controlled the output; naming a task, saying that a tool was used sparingly, or accepting responsibility alone does not qualify. Numerical agreement between models and validation reproduced from cited studies do not count as human oversight. Separate labels record the human-oversight and limitation components, responsibility language, whether an individual operator is named, whether the wording is compatible with a publisher template, and whether reported writing use is confined to surface editing, proofreading, translation, or typesetting. The labels were proposed by the reviewing model from the retrieved contexts, every positive oversight or restriction proposal received a second source-context check, and the historical primary-session review adjudicated 34 cases against the rendered pages. The first-pass and adjudicated protective labels agree for 1,049 of the 1,167 disclosures. No human validation of these labels has been completed, and the checks share one manual, so shared systematic errors remain possible. Section~\ref{sec:wording-appendix} reports the resulting comparisons.

Colons and em/en dashes are counted in the same selected material as the writing scores and expressed per 1,000 words. The pooled high-rate thresholds are 8.905 for colons and 9.267 for dashes, each fixed at the mean plus two standard deviations in a historical 2015--2021 reference sample. The raw dash measure includes numeric ranges and decorative rules; selected material can also retain tables and dash-shaped negative signs. The cleaned measure removes numeric ranges and repeated glyph runs. It remains a mixture of exposition, content, and typesetting. Source-specific historical flags and alternative constructions enter the complete supporting result family in Appendix~\ref{sec:punctuation}.

Researcher identities principally use OpenAlex works, identifiers, and corroborated author rosters. For recent papers, a flagged extension reuses an already linked identity where a full byline name is unique in the linked population. Ambiguous names remain unmatched. Table~\ref{tab:identity} excludes these name-only extensions. Byline counts are measured independently of identity linkage. An indexed identity can contain attribution error, and the included linked researchers need not cover every printed author. Equation~\ref{eq:empirical} gives each paper total weight one across its included researchers.

Additional outcomes distinguish comments/readership, solo authorship, and solo authorship without assistance, and transform named counts using $\log(1+\text{count})$ or a cap at eight names, the fixed 99th percentile of the pre-2019 reference count distribution. Eight documented attributes describe formal theory, computational execution, existing-data use, original information acquisition, remote implementation, physical implementation, public data access, and restricted access. All eight must be resolved for their sum to be observed. Non-documentation does not establish absence. Appendix~\ref{sec:attributes} examines components whose evidence can overlap with disclosure coding.

\subsection{Samples and overlap}
\begin{table}[htbp]\centering\caption{From observed measures to within-researcher samples}\label{tab:flow}
{\small\renewcommand{\arraystretch}{1.12}\setlength{\tabcolsep}{4pt}\begin{tabular*}{\linewidth}{@{\extracolsep{\fill}}lrrrr@{}}
\toprule
Outcome & \shortstack{Observed\\measures} & \shortstack{Linked\\complete} & \shortstack{Final\\papers} & \shortstack{Disclosure\\switchers}\\\midrule
Any assistant (pp) & 11,506 & 9,715 & 7,243 & 535\\
Named assistants & 11,505 & 9,714 & 7,242 & 535\\
Authors & 13,713 & 11,516 & 8,502 & 633\\
Pages & 13,713 & 11,516 & 8,502 & 633\\
Documented attributes & 11,912 & 10,129 & 7,251 & 487\\
\bottomrule\end{tabular*}}\tabnote{All outcomes begin with 16,721 timing-supported post-2022 papers. Observed measures require the outcome, both indicators and controls. Linked complete cases additionally require usable researcher identities. Final papers survive recurrence and fixed-effect pruning. Switchers count included researchers with both disclosure values on different papers; this is not the number of treated papers.}\end{table}

\begin{table}[htbp]\centering\caption{Post-2022 paper characteristics}\label{tab:descriptive}
{\small\renewcommand{\arraystretch}{1.12}\setlength{\tabcolsep}{4pt}\begin{tabular*}{\linewidth}{@{\extracolsep{\fill}}lrrrr@{}}
\toprule
Measure & Papers & Mean & SD & Median\\\midrule
Authors & 16,720 & 2.890 & 3.118 & 3.000\\
Any assistant & 13,784 & 0.245 & 0.430 & 0.000\\
Named assistants & 13,783 & 0.697 & 1.989 & 0.000\\
Pages & 16,721 & 48.903 & 25.128 & 46.000\\
Writing flag & 14,017 & 0.084 & 0.277 & 0.000\\
Disclosed use & 16,301 & 0.072 & 0.258 & 0.000\\
Solo author & 16,720 & 0.166 & 0.372 & 0.000\\
Solo, no assistance & 13,543 & 0.111 & 0.315 & 0.000\\
\bottomrule\end{tabular*}}\tabnote{One observation per timing-supported paper, January 2023--October 2026. Each measure uses its observed denominator before requiring linked recurring researchers. Binary measures are fractions. These samples differ from the regression samples.}\end{table}

The 13,713 post-2022 papers observed on both principal indicators contain 992 disclosures, 1,139 writing flags and 137 joint positives. These are jointly observed descriptive counts before the recurrence and outcome restrictions used in the regressions. Of these papers, 11,719 carry neither recorded indicator, 855 disclose without a writing flag and 1,002 carry a writing flag without disclosure. The overlap percentages in the main text condition on the relevant positive-indicator denominator, not on all papers.

For a partition into cells $c$, the independence reference is $\sum_c n_{Dc}n_{Wc}/n_c$. Its expected joint count is 82.40 when pooled, 107.70 within source-year cells and 100.82 within source-quarter cells. The observed-to-expected ratios are respectively 1.66, 1.27 and 1.36. These are descriptions of recorded indicators on the same papers, not estimates of the correlation between unobserved uses. They allow for common variation in indicator prevalence without attributing low overlap to a single cause.

\subsection{Measurement quality}\label{sec:measurement-quality}
Disclosure measurement combines candidate retrieval, document-level review, and literal source adjudication. The initial historical classification used GPT-5.6 Luna. A broader corpus-wide screen then retrieved passages for contextual review with GPT-6 Astra and tools disabled. Reviews distinguish author assistance from topical mentions, explicit non-use, and unresolved statements. A positive label requires a literal source passage; no statement in assessable text is coded negative, and missing or ambiguous evidence remains unresolved. Historical labels on retained documents are preserved. New-source review covers 4,785 source hashes and 1,079 document decisions, with positive oversight and restriction proposals checked against their surrounding passages. These procedures are model-assisted documentary measurement, not independent human validation.

The selected frame contains 1,729 positive disclosure labels, 53,617 zeros, and 1,239 missing values. Review and staging receipts support the requested GPT-6 Astra model and disabled tools. Component-review events do not independently identify the runtime model: their model attribution rests on the component protocol, unchanged command source, and coordinator record rather than a separate per-batch model receipt. This limits the provenance of the component classifications. Lexically selected candidate review does not establish corpus-wide recall. A stratified historical draw contains only 36 resolved reference labels among the 584 records that match the retained frame.

Task classifications apply to the whole document. A statement about editing does not make a paper writing-only if another passage reports AI-assisted research operations. Positive human-oversight labels require evidence of human checking; numerical agreement, validation reproduced from cited studies, and model-to-model agreement do not qualify. Missing statements and incomplete source text can still produce measurement error. Source evidence, decision ledgers, coding rules, and full outputs accompany the replication documentation.

\subsection{Reference lists and institutional affiliation}\label{sec:references-affiliations}
This section reports the within-researcher reference-list and university-affiliation comparisons summarised in Section 5.5, alongside bibliography coverage and conditioning checks across all 24 series. In the opening summary table, disclosing papers cite literature with a median age 1.24 years younger (SE 0.31; Romano--Wolf \(p = .0005\); Holm \(p = .0003\)). Bibliography length differs by +0.044 \(\log_2\) references (SE 0.041; Holm \(p = .57\)), and the three other tests in the four-test reference family all have Romano--Wolf \(p\) above .14. Among disclosing papers, 19.3\% mention `artificial intelligence' or `AI' in the title or abstract against 2.5\% of other papers; excluding them lowers the age gap to 0.54 years (SE 0.32). Adding paper-level university-tier shares leaves the disclosure estimates for any assistant and total pages unchanged at 6.63 percentage points and 5.41 pages (matched baselines 6.60 and 5.35).

\begin{table}[htbp]\centering\caption{Reference lists and institutional composition}\label{tab:references_main}
{\small\renewcommand{\arraystretch}{1.10}\setlength{\tabcolsep}{3pt}
\begin{tabular*}{\linewidth}{@{\extracolsep{\fill}}lrr@{}}
\toprule
 & \shortstack{Median age\\(years)} & \shortstack{Reference count\\(log base 2)}\\
\midrule
Writing flag & \begin{tabular}[t]{@{}c@{}}-0.539\\(0.267)\end{tabular} & \begin{tabular}[t]{@{}c@{}}0.040\\(0.042)\end{tabular}\\
Holm probability & 0.131 & 0.565\\
Disclosed use & \begin{tabular}[t]{@{}c@{}}-1.239\\(0.312)\end{tabular} & \begin{tabular}[t]{@{}c@{}}0.044\\(0.041)\end{tabular}\\
Holm probability & 0.000 & 0.565\\
Papers & 8,125 & 8,392\\
Disclosed papers & 512 & 534\\
Researchers & 5,084 & 5,218\\
Sources & 24 & 24\\
\multicolumn{3}{@{}l}{University composition: disclosure coefficient}\\
 & Any assistant (pp) & Pages\\
Matched baseline & \begin{tabular}[t]{@{}c@{}}6.599\\(2.606)\end{tabular} & \begin{tabular}[t]{@{}c@{}}5.346\\(1.274)\end{tabular}\\
University tier shares & \begin{tabular}[t]{@{}c@{}}6.631\\(2.606)\end{tabular} & \begin{tabular}[t]{@{}c@{}}5.410\\(1.270)\end{tabular}\\
Papers & 7,243 & 8,502\\
\bottomrule
\end{tabular*}}\tabnote{Joint indicators and the specification in equation~\ref{eq:empirical}. Upper panel: counts use log base two of one plus the number of references. Eligible reference measures determine the corresponding fit; missing measures are withheld. Lower panel: paper-level shares of authors in THE 2026 university tiers, with unresolved slots explicit. Each baseline uses identical papers to its augmented comparison. Parentheses show researcher/paper-clustered standard errors. Source and shared-author sensitivities are reported separately.}\end{table}

Reference measurement covers all 24 series and uses the exact PDFs for the union of the ten fitted samples and the pre-2022 historical samples: 8,502 contemporary and 6,563 historical papers in the frozen measurement frame. Counts include distinct entries, including undated and non-journal items. Age uses the verified version year where available and the printed-date assumption otherwise. It requires at least five dated entries, dates for 80\% of entries, and no unresolved future citation years. Incomplete lists and known extraction errors remain missing. Counts are available for 8,416 contemporary papers and ages for 8,198; Table~\ref{tab:reference_coverage} reports coverage by source and period.

Independent model-assisted text readings assess bibliography extraction in stratified documentary samples. Count agreement permits two entries or 5\% of the independently enumerated count, whichever is larger; age agreement requires the same eligibility decision and a median within one year. A historical 82-paper diagnostic sample gives count agreement of 21/24 for NBER, 4/4 for arXiv and 49/54 for other series; age agreement is 23/24, 4/4 and 52/54. A separate source-period-stratified sample gives weighted count/age agreement of NBER (8 papers; 100.0\%/100.0\%); arXiv (2 papers; 50.0\%/100.0\%); other series (14 papers; 85.7\%/100.0\%); Bank of England (30 papers; 97.3\%/100.0\%); Boston Fed (30 papers; 93.3\%/95.8\%). The samples cover distinct source-document batches. Their agreement rates are diagnostic and are not pooled into a population accuracy estimate.

Historical rendered-page checks include a 27-paper draw from series other than NBER and arXiv, with 98.1\% weighted count agreement (25/27) and 94.9\% age agreement (24/27). A separate draw completes 6 of 10 selected readings, with count agreement on 6/6 and age agreement on 6/6. Bibliographies exceeding the eight-page review limit are recorded as unavailable for this check. These fallible readings reveal merged entries, appendix material counted as references, and disputed citation years. Affected measures are withheld, without an accuracy threshold for admitting a series.

The additional reference audit gives weighted count agreement of 87.34\% across 86 selected papers and age agreement of 95.14\% across 85 eligible papers. Missing documentary readings remain in these denominators. The rendered-reference draw completes three of ten selected readings; all three agree on count and age, and seven are unavailable. These rates are reported separately from the historical samples. Known failed measures remain withheld.

\begin{table}[htbp]\centering\caption{Reference coverage by source and period}\label{tab:reference_coverage}
{\small\renewcommand{\arraystretch}{1.10}\setlength{\tabcolsep}{3pt}
\begin{tabular*}{\linewidth}{@{\extracolsep{\fill}}lrrrrrr@{}}
\toprule
 & \multicolumn{3}{c}{Historical} & \multicolumn{3}{c}{Contemporary}\\
Source & Target & Count & Age & Target & Count & Age\\
\midrule
arXiv economics & 534 & 521 & 481 & 1,872 & 1,847 & 1,775\\
Bank of Canada & 106 & 105 & 102 & 60 & 58 & 59\\
BIS & 99 & 97 & 92 & 95 & 92 & 93\\
Bank of England & 77 & 76 & 74 & 54 & 52 & 51\\
Warwick CAGE & 90 & 88 & 86 & 75 & 74 & 71\\
Cambridge & 106 & 105 & 99 & 86 & 86 & 82\\
US Census CES & 20 & 20 & 19 & 74 & 73 & 73\\
LSE CEP & 117 & 115 & 112 & 72 & 72 & 72\\
CESifo & 1,163 & 1,154 & 1,122 & 1,516 & 1,508 & 1,486\\
Osaka ISER & 66 & 65 & 62 & 43 & 42 & 42\\
ANU CAMA & 116 & 114 & 109 & 81 & 81 & 81\\
Boston Fed & 22 & 19 & 20 & 35 & 31 & 31\\
San Francisco Fed & 30 & 29 & 29 & 46 & 44 & 45\\
Federal Reserve Board & 119 & 117 & 115 & 138 & 138 & 135\\
Chicago Fed & 23 & 23 & 23 & 30 & 28 & 29\\
Institute for Fiscal Studies & 46 & 46 & 45 & 85 & 84 & 84\\
NBER & 2,724 & 2,701 & 2,613 & 3,123 & 3,101 & 3,018\\
New York Fed & 55 & 53 & 50 & 45 & 45 & 44\\
Philadelphia Fed & 62 & 61 & 59 & 60 & 59 & 56\\
RIETI & 90 & 90 & 88 & 31 & 29 & 30\\
Tinbergen & 102 & 101 & 99 & 70 & 70 & 69\\
Toulouse & 120 & 115 & 111 & 158 & 157 & 153\\
Pompeu Fabra & 46 & 44 & 40 & 9 & 9 & 8\\
World Bank & 630 & 627 & 594 & 644 & 636 & 611\\
\bottomrule
\end{tabular*}}\tabnote{Each row is a source series. Historical and contemporary target counts are the frozen measurement-frame populations; count and age columns report the corresponding paper-level measures available for analysis. Missing measures are withheld, not imputed. Age requires at least five dated references, dates for at least 80\% of entries and no unresolved future citation years.}\end{table}

Across the 24 series, median reference counts are 43 in the historical sample and 47 in the contemporary sample; median ages are 7 and 8 years, respectively. These are descriptions of selected papers, not a panel of unchanged projects. Regression counts use $\log_2(1+\text{references})$ and age enters the human-input comparisons per five years. Table~\ref{tab:reference_controls} retains identical-sample baselines. The four reference-outcome coefficients and fifteen reference-conditioned coefficients form separate Holm families. In the strictly dated sample, which contains NBER and arXiv papers, the disclosure coefficient for age is $-0.77$ years (standard error 0.61; 2,213 papers). The single-source disclosure/age estimates are $-1.04$ years (0.53; 2,854 papers) for NBER and $-0.64$ (0.73; 1,604 papers) for arXiv. Full single-source, timing, and shared-author sensitivities accompany the results.

Source-score tests give Holm-adjusted probabilities of 0.002 for disclosure in the age regression and 0.360 in the count regression, with about 3.6 effective sources for age. Shared-author covariance leaves their standard errors close to those in Table~\ref{tab:references_main}. Thus coverage across many series does not make the identifying information evenly distributed across them.

\begin{table}[htbp]\centering\caption{Research inputs conditional on reference characteristics}\label{tab:reference_controls}
{\small\renewcommand{\arraystretch}{1.10}\setlength{\tabcolsep}{3pt}
\begin{tabular*}{\linewidth}{@{\extracolsep{\fill}}lrrrr@{}}
\toprule
 & \shortstack{Any assistant\\(pp)} & \shortstack{Named\\assistants} & Authors & Pages\\
\midrule
Writing: matched baseline & \begin{tabular}[t]{@{}c@{}}-3.459\\(2.327)\end{tabular} & \begin{tabular}[t]{@{}c@{}}-0.120\\(0.117)\end{tabular} & \begin{tabular}[t]{@{}c@{}}0.012\\(0.063)\end{tabular} & \begin{tabular}[t]{@{}c@{}}-0.768\\(1.152)\end{tabular}\\
Writing: reference measures & \begin{tabular}[t]{@{}c@{}}-3.834\\(2.344)\end{tabular} & \begin{tabular}[t]{@{}c@{}}-0.127\\(0.119)\end{tabular} & \begin{tabular}[t]{@{}c@{}}-0.002\\(0.063)\end{tabular} & \begin{tabular}[t]{@{}c@{}}-0.743\\(1.149)\end{tabular}\\
Disclosure: matched baseline & \begin{tabular}[t]{@{}c@{}}8.115\\(2.712)\end{tabular} & \begin{tabular}[t]{@{}c@{}}0.618\\(0.177)\end{tabular} & \begin{tabular}[t]{@{}c@{}}0.219\\(0.097)\end{tabular} & \begin{tabular}[t]{@{}c@{}}5.357\\(1.302)\end{tabular}\\
Disclosure: reference measures & \begin{tabular}[t]{@{}c@{}}7.692\\(2.696)\end{tabular} & \begin{tabular}[t]{@{}c@{}}0.616\\(0.177)\end{tabular} & \begin{tabular}[t]{@{}c@{}}0.187\\(0.096)\end{tabular} & \begin{tabular}[t]{@{}c@{}}5.414\\(1.301)\end{tabular}\\
Median age / 5 & \begin{tabular}[t]{@{}c@{}}-1.447\\(0.579)\end{tabular} & \begin{tabular}[t]{@{}c@{}}0.014\\(0.033)\end{tabular} & \begin{tabular}[t]{@{}c@{}}-0.127\\(0.015)\end{tabular} & \begin{tabular}[t]{@{}c@{}}0.230\\(0.328)\end{tabular}\\
\begin{tabular}[t]{@{}c@{}}Reference count\\(log base 2)\end{tabular} & \begin{tabular}[t]{@{}c@{}}4.098\\(0.865)\end{tabular} & \begin{tabular}[t]{@{}c@{}}0.178\\(0.042)\end{tabular} & \begin{tabular}[t]{@{}c@{}}0.011\\(0.027)\end{tabular} & --\\
Papers & 6,923 & 6,923 & 8,099 & 8,125\\
Disclosed papers & 444 & 444 & 512 & 512\\
\bottomrule
\end{tabular*}}\tabnote{Each baseline uses exactly the rows, papers and paper-total weights of its corresponding augmented specification. Both AI indicators, original fixed effects and prose controls remain included. Researcher/paper-clustered standard errors are shown in parentheses; source and shared-author diagnostics accompany these estimates. Counts use log base two of one plus the number of references. Reference count is omitted for pages because bibliographies contribute directly to document length. The fifteen writing, disclosure, age and count coefficients form one Holm family.}\end{table}

University affiliation comes from the measured paper's author-information pages, with same-year, own-DOI preprint metadata as a restricted fallback. Each author receives the best explicitly affiliated university's category in the \href{https://www.timeshighereducation.com/world-university-rankings/2026/world-ranking}{THE overall 2026 ranking}. This fixed edition contains 2,191 ranked institutions and 927 reporters. Categories are top 100, 101--200, 201--500, 501+, known unranked university, non-university, and unresolved; research-network membership does not substitute for an employer. This is a retrospective institutional classification, not a measure of an economist's individual quality.

The mapping resolves 84.6\% of 26,242 printed author slots. A historical 100-paper probability sample yields 76.4\% weighted tier agreement against independent model-assisted front-matter readings and a separate ranking match (paper-clustered 95\% interval 67.5--85.4\%). A separate blinded rendered-page sample contains 18 papers and 56 author slots. Of these, 12 paper readings were completed; 6 could not be completed within the fixed measurement budget. Counting unavailable readings and unresolved decisions as disagreement gives 61.1\% weighted tier agreement using the same frozen ranking crosswalk. These readings do not independently validate the crosswalk, and the diagnostic samples are not pooled. In the authorship sample, unresolved slots make up 16.4\% of printed slots on disclosing papers and 15.4\% on other papers, a difference of 1.0 percentage points that falls to 0.3 after source standardisation. The additional affiliation audit gives 61.7\% weighted tier agreement across 105 slots in 28 selected papers. Readings were completed for 24 papers; the four unavailable readings count as disagreements. These documentary checks do not establish human validation, and the samples are not pooled. Mapping errors limit institutional comparisons.

Tier shares divide by all printed authors, retaining unresolved slots and omitting non-university affiliation. Their author-count comparison conditions on its own denominator. Interactions use each row researcher's top-100 status and its identified main effect, excluding unknown statuses and recalculating recurrence and weights; all other resolved categories form the comparison group. The eight interactions have Holm-adjusted probabilities of at least 0.332, without establishing institutional equality. These models contain 6,129--7,045 papers. Each side has at least 195 disclosed papers from 13 sources and 176 researchers with within-status disclosure variation. Writing-flag support is at least 153 positive papers from 12 sources and 152 researchers with within-status variation. The eight tier-conditioned coefficients and eight interactions form separate Holm families. Table~\ref{tab:university} reports estimates; complete-roster, NBER-omitted, and combined reference/rank comparisons accompany the results files.

\begin{table}[htbp]\centering\caption{University affiliation and research inputs}\label{tab:university}
{\small\renewcommand{\arraystretch}{1.10}\setlength{\tabcolsep}{3pt}
\begin{tabular*}{\linewidth}{@{\extracolsep{\fill}}lrrrr@{}}
\toprule
 & \shortstack{Any assistant\\(pp)} & \shortstack{Named\\assistants} & Authors & Pages\\
\midrule
Writing: matched baseline & \begin{tabular}[t]{@{}c@{}}-3.206\\(2.251)\end{tabular} & \begin{tabular}[t]{@{}c@{}}-0.131\\(0.111)\end{tabular} & \begin{tabular}[t]{@{}c@{}}0.011\\(0.061)\end{tabular} & \begin{tabular}[t]{@{}c@{}}-0.693\\(1.100)\end{tabular}\\
Writing: tier shares & \begin{tabular}[t]{@{}c@{}}-3.290\\(2.240)\end{tabular} & \begin{tabular}[t]{@{}c@{}}-0.134\\(0.110)\end{tabular} & \begin{tabular}[t]{@{}c@{}}0.004\\(0.061)\end{tabular} & \begin{tabular}[t]{@{}c@{}}-0.667\\(1.098)\end{tabular}\\
Disclosure: matched baseline & \begin{tabular}[t]{@{}c@{}}6.599\\(2.606)\end{tabular} & \begin{tabular}[t]{@{}c@{}}0.548\\(0.168)\end{tabular} & \begin{tabular}[t]{@{}c@{}}0.240\\(0.093)\end{tabular} & \begin{tabular}[t]{@{}c@{}}5.346\\(1.274)\end{tabular}\\
Disclosure: tier shares & \begin{tabular}[t]{@{}c@{}}6.631\\(2.606)\end{tabular} & \begin{tabular}[t]{@{}c@{}}0.551\\(0.168)\end{tabular} & \begin{tabular}[t]{@{}c@{}}0.233\\(0.093)\end{tabular} & \begin{tabular}[t]{@{}c@{}}5.410\\(1.270)\end{tabular}\\
Writing $\times$ own top 100 & \begin{tabular}[t]{@{}c@{}}-1.169\\(4.854)\end{tabular} & \begin{tabular}[t]{@{}c@{}}0.045\\(0.264)\end{tabular} & \begin{tabular}[t]{@{}c@{}}-0.133\\(0.138)\end{tabular} & \begin{tabular}[t]{@{}c@{}}4.808\\(2.358)\end{tabular}\\
Disclosure $\times$ own top 100 & \begin{tabular}[t]{@{}c@{}}-2.802\\(5.144)\end{tabular} & \begin{tabular}[t]{@{}c@{}}-0.124\\(0.317)\end{tabular} & \begin{tabular}[t]{@{}c@{}}0.137\\(0.191)\end{tabular} & \begin{tabular}[t]{@{}c@{}}0.901\\(2.514)\end{tabular}\\
Papers: shares & 7,243 & 7,242 & 8,502 & 8,502\\
Papers: interactions & 6,130 & 6,129 & 7,045 & 7,045\\
\bottomrule
\end{tabular*}}\tabnote{THE overall 2026 ranking applied retrospectively to paper-specific affiliations. Tier shares divide by all printed authors, with unresolved slots explicit; their authors column is mechanically conditioned on the outcome denominator. Interaction coefficients use the row researcher\textquotesingle s own paper-specific top-100 status, excluding unresolved statuses and recomputing recurrence and paper weights. Parentheses show researcher/paper-clustered standard errors. Eight AI coefficients with tier controls and eight institutional interactions form separate Holm families. These are descriptive institutional comparisons.}\end{table}

\FloatBarrier

The disclosure-on-team-size comparison starts from the 8,502-paper authors sample. Raw disclosure shares for one, two, three, four, and five-or-more authors are 10.2\%, 5.3\%, 5.4\%, 6.5\% and 8.5\%, respectively; denominators are 783, 2,200, 2,800, 1,735 and 984. The conditional log-base-two author-count slope is 1.36 percentage points (standard error 0.60). Disclosure information for that slope is concentrated in 4.1 effective sources, and its source-clustered interval spans zero. Giving eligible sources equal weight changes the slope from 1.21 to 2.17 percentage points on the same 9-source, 7,549-paper subset. On the 8,099-paper reference-and-rank subset, the matched slope is 1.18 percentage points (standard error 0.60), compared with 0.95 (0.60) after conditioning. These three slopes form one Holm family; none identifies author preferences.

\FloatBarrier\clearpage\setcounter{table}{0}\setcounter{figure}{0}
\section{Supporting empirical results}\label{app:results}
\subsection{Full outcomes and uncertainty}\label{sec:inference}
Table~\ref{tab:full} gives all ten outcomes for the joint writing/disclosure model and the pooled comparison on identical samples. The two indicators across the ten within-researcher outcomes form a twenty-test family; Table~\ref{tab:probabilities} reports Holm-adjusted probabilities separately for researcher/paper clustering and source-score inference. Additional restrictions are descriptive sensitivities.

The source calculation is a null-restricted wild score test. Nuisance coefficients and fixed effects are estimated with the tested indicator's coefficient set to zero; source-level restricted scores receive Rademacher signs in 9,999 fixed-seed draws. It is an unstudentised score calculation, not a bootstrap-$t$ interval. With disclosure information concentrated in about 3.3 effective sources, few influential signs can dominate its distribution. The calculation also does not automatically cover coauthor dependence across outlets, so small adjusted probabilities are conditional summaries rather than generally reliable staffing tests.

For shared authors, weighted, fixed-effect-partialled scores are first summed to each paper. The covariance sensitivity includes the outer products for the same paper and every pair sharing a reliably linked author, counting each ordered pair once even when several authors overlap. It allows covariance between whole papers, beyond rows bearing the same researcher label, but assumes no covariance outside linked neighbours. Incomplete linkage and broader source shocks can violate this assumption. Paper-only and overlap calculations omit small-sample corrections; the main researcher/paper standard errors use the software defaults.

In the assistance-incidence sample, linked identities cover 92.5\% of byline slots. The largest connected component contains 60.8\% of paper weight, and 1,844 linked authors span sources. Strict work-corroborated links cover 90.6\% of slots, with 58.2\% in the largest component. This extensive connection and incomplete coverage limit both one-hop covariance and any interpretation that assumes independent outlets. Table~\ref{tab:dependence} reports the resulting sensitivities alongside the concentration of disclosure information.

\begin{table}[htbp]\centering\caption{Family-adjusted probabilities for the complete outcome set}\label{tab:probabilities}
{\small\renewcommand{\arraystretch}{1.12}\setlength{\tabcolsep}{4pt}\begin{tabular*}{\linewidth}{@{\extracolsep{\fill}}lrrrr@{}}
\toprule
Outcome & \shortstack{Writing\\row clusters} & \shortstack{Disclosure\\row clusters} & \shortstack{Writing\\source score} & \shortstack{Disclosure\\source score}\\\midrule
Any assistant (pp) & 1.000 & 0.170 & 1.000 & 1.000\\
Named assistants & 1.000 & 0.019 & 1.000 & 0.036\\
Authors & 1.000 & 0.163 & 1.000 & 1.000\\
Pages & 1.000 & $<0.001$ & 1.000 & 0.028\\
Documented attributes & 1.000 & 0.772 & 1.000 & 1.000\\
Solo author (pp) & 1.000 & 1.000 & 1.000 & 1.000\\
Solo, no assistance (pp) & 1.000 & 1.000 & 1.000 & 1.000\\
Any reader (pp) & 1.000 & 1.000 & 1.000 & 1.000\\
$\log(1+\mathrm{assistants})$ & 1.000 & 0.004 & 1.000 & 0.537\\
Capped assistants & 1.000 & 0.002 & 1.000 & 0.049\\
\bottomrule\end{tabular*}}\tabnote{Holm adjustment across two indicators and ten outcomes, separately for each inference method. Row clusters means researcher/paper clustering. Source scores use 9,999 null-restricted Rademacher draws. Few effective sources and incomplete coverage of shared-author dependence prevent treating these probabilities as generally robust inference; a small value does not remove that limitation. These are exploratory comparisons.}\end{table}

\begin{table}[htbp]\centering\caption{Pooled and within-researcher comparisons}\label{tab:full}
{\footnotesize\renewcommand{\arraystretch}{1.12}\setlength{\tabcolsep}{2pt}\begin{tabular*}{\linewidth}{@{\extracolsep{\fill}}>{\raggedright\arraybackslash}p{1.25in}rrrrrr@{}}
\toprule
Outcome & \shortstack{Writing\\pooled} & \shortstack{Writing\\within} & \shortstack{Disclosure\\pooled} & \shortstack{Disclosure\\within} & Papers & \shortstack{RW $p$\\W / D}\\\midrule
Any assistant (pp) & \begin{tabular}[t]{@{}c@{}}-8.870\\(2.142)\end{tabular} & \begin{tabular}[t]{@{}c@{}}-3.206\\(2.251)\end{tabular} & \begin{tabular}[t]{@{}c@{}}8.949\\(2.422)\end{tabular} & \begin{tabular}[t]{@{}c@{}}6.599\\(2.606)\end{tabular} & 7,243 & \begin{tabular}[t]{@{}c@{}}.8185 / .157\end{tabular}\\
Named assistants & \begin{tabular}[t]{@{}c@{}}-0.349\\(0.085)\end{tabular} & \begin{tabular}[t]{@{}c@{}}-0.131\\(0.111)\end{tabular} & \begin{tabular}[t]{@{}c@{}}0.585\\(0.148)\end{tabular} & \begin{tabular}[t]{@{}c@{}}0.548\\(0.168)\end{tabular} & 7,242 & \begin{tabular}[t]{@{}c@{}}.898 / .02\end{tabular}\\
Authors & \begin{tabular}[t]{@{}c@{}}-0.010\\(0.072)\end{tabular} & \begin{tabular}[t]{@{}c@{}}0.011\\(0.061)\end{tabular} & \begin{tabular}[t]{@{}c@{}}0.207\\(0.100)\end{tabular} & \begin{tabular}[t]{@{}c@{}}0.240\\(0.093)\end{tabular} & 8,502 & \begin{tabular}[t]{@{}c@{}}.9975 / .147\end{tabular}\\
Pages & \begin{tabular}[t]{@{}c@{}}-2.454\\(0.971)\end{tabular} & \begin{tabular}[t]{@{}c@{}}-0.693\\(1.100)\end{tabular} & \begin{tabular}[t]{@{}c@{}}4.973\\(1.212)\end{tabular} & \begin{tabular}[t]{@{}c@{}}5.346\\(1.274)\end{tabular} & 8,502 & \begin{tabular}[t]{@{}c@{}}.9935 / .0005\end{tabular}\\
Documented attributes & \begin{tabular}[t]{@{}c@{}}-0.165\\(0.062)\end{tabular} & \begin{tabular}[t]{@{}c@{}}-0.083\\(0.064)\end{tabular} & \begin{tabular}[t]{@{}c@{}}0.249\\(0.072)\end{tabular} & \begin{tabular}[t]{@{}c@{}}0.133\\(0.069)\end{tabular} & 7,251 & \begin{tabular}[t]{@{}c@{}}.862 / .488\end{tabular}\\
Solo author (pp) & \begin{tabular}[t]{@{}c@{}}3.319\\(1.712)\end{tabular} & \begin{tabular}[t]{@{}c@{}}1.331\\(1.282)\end{tabular} & \begin{tabular}[t]{@{}c@{}}3.994\\(2.302)\end{tabular} & \begin{tabular}[t]{@{}c@{}}0.359\\(1.658)\end{tabular} & 8,502 & \begin{tabular}[t]{@{}c@{}}.921 / .9975\end{tabular}\\
Solo, no assistance (pp) & \begin{tabular}[t]{@{}c@{}}1.865\\(1.463)\end{tabular} & \begin{tabular}[t]{@{}c@{}}1.019\\(1.295)\end{tabular} & \begin{tabular}[t]{@{}c@{}}1.160\\(1.606)\end{tabular} & \begin{tabular}[t]{@{}c@{}}-0.144\\(1.373)\end{tabular} & 7,099 & \begin{tabular}[t]{@{}c@{}}.979 / .9975\end{tabular}\\
Any reader (pp) & \begin{tabular}[t]{@{}c@{}}-4.254\\(2.618)\end{tabular} & \begin{tabular}[t]{@{}c@{}}-0.823\\(2.686)\end{tabular} & \begin{tabular}[t]{@{}c@{}}0.149\\(2.412)\end{tabular} & \begin{tabular}[t]{@{}c@{}}-1.178\\(2.577)\end{tabular} & 7,136 & \begin{tabular}[t]{@{}c@{}}.9975 / .9935\end{tabular}\\
$\log(1+\mathrm{assistants})$ & \begin{tabular}[t]{@{}c@{}}-0.121\\(0.026)\end{tabular} & \begin{tabular}[t]{@{}c@{}}-0.045\\(0.029)\end{tabular} & \begin{tabular}[t]{@{}c@{}}0.173\\(0.039)\end{tabular} & \begin{tabular}[t]{@{}c@{}}0.145\\(0.039)\end{tabular} & 7,242 & \begin{tabular}[t]{@{}c@{}}.7545 / .0035\end{tabular}\\
Capped assistants & \begin{tabular}[t]{@{}c@{}}-0.306\\(0.065)\end{tabular} & \begin{tabular}[t]{@{}c@{}}-0.125\\(0.075)\end{tabular} & \begin{tabular}[t]{@{}c@{}}0.507\\(0.111)\end{tabular} & \begin{tabular}[t]{@{}c@{}}0.426\\(0.110)\end{tabular} & 7,242 & \begin{tabular}[t]{@{}c@{}}.6495 / .0015\end{tabular}\\
\bottomrule\end{tabular*}}\tabnote{RW reports Romano--Wolf probabilities for the within-researcher writing (W) and disclosure (D) tests, using 1,999 researcher-level wild-bootstrap draws and two-way studentisation across the twenty-test family. Each outcome uses identical papers in its pooled and within-researcher comparisons. Pooled specifications omit researcher effects; other controls and paper-total weighting follow equation~\ref{eq:empirical}. Standard errors in parentheses cluster by researcher and paper and are conditional on that structure. Binary outcomes use percentage points. All twenty within-researcher indicator/outcome tests enter the family in Table~\ref{tab:probabilities}.}\end{table}

\begin{table}[htbp]\centering\caption{Source concentration and shared-author uncertainty}\label{tab:dependence}
{\small\renewcommand{\arraystretch}{1.12}\setlength{\tabcolsep}{4pt}\begin{tabular*}{\linewidth}{@{\extracolsep{\fill}}lrrrrr@{}}
\toprule
Outcome & \shortstack{Effective\\sources} & \shortstack{Largest\\share (\%)} & \shortstack{Paper\\SE} & \shortstack{All-link\\overlap SE} & \shortstack{Strict-link\\overlap SE}\\\midrule
Any assistant (pp) & 3.34 & 45.6 & 2.177 & 2.693 & 2.635\\
Named assistants & 3.34 & 45.6 & 0.143 & 0.182 & 0.179\\
Authors & 3.60 & 38.7 & 0.080 & 0.100 & 0.098\\
Pages & 3.60 & 38.7 & 1.029 & 1.354 & 1.324\\
\bottomrule\end{tabular*}}\tabnote{Effective sources is the inverse sum of squared source shares of residualised disclosure information. Overlap covariance sums whole-paper scores across each linked shared-author pair once. Strict links exclude name-only extensions. Paper-only and overlap calculations omit small-sample corrections; the main standard errors use the software corrections. Incomplete linkage and source shocks limit these sensitivities; they are not generally dependence-robust standard errors.}\end{table}

\FloatBarrier

\subsection{Documentation, source composition, and sample restrictions}\label{sec:sensitivity}
The documentation comparisons use matched samples: each baseline retains exactly the papers in its corresponding conditioned specification (Table~\ref{tab:documentation}). Both include the same joint indicators and fixed effects. Full-text length conditioning is not reported for page count because it would condition document length on a closely related outcome.

Large-list and strict-timing restrictions appear in Table~\ref{tab:restrictions}. Each refit re-evaluates eligibility and fixed-effect support and recomputes paper-total weights. These restrictions change the population being described. In particular, positive point estimates in the small strict-date sample do not establish temporal comparability of the larger printed-date sample.

The equal-source comparison retains sources with at least five disclosed and twenty no-statement papers in the final outcome-specific sample. Its ordinary-weight and equal-weight rows use identical papers (Table~\ref{tab:balanced}). For assistant incidence, the eight sources are arXiv economics, Warwick CAGE, Cambridge, CESifo, Federal Reserve Board, Institute for Fiscal Studies, NBER, and World Bank. Renewed recurrence and fixed-effect restrictions leave 6,387 papers for both weighting schemes. Equal weighting gives each source the same total weight, not the same residual information. The equal-source incidence interval spans negative and positive differences.

Table~\ref{tab:omissions} reports the two largest-source omissions and the range over every single-source omission. Tables~\ref{tab:balanced} and \ref{tab:identity} distinguish source reweighting from stricter identity evidence. Complete source-by-outcome support, omissions, uncertainty, and weights accompany the result files; the ranges are not selected confidence bounds.

\begin{table}[htbp]\centering\caption{Disclosure and documentation on matched samples}\label{tab:documentation}
{\small\renewcommand{\arraystretch}{1.12}\setlength{\tabcolsep}{4pt}\begin{tabular*}{\linewidth}{@{\extracolsep{\fill}}lrrrr@{}}
\toprule
 & \shortstack{Any assistant\\(pp)} & \shortstack{Named\\assistants} & Authors & Pages\\\midrule
Acknowledgement baseline & \begin{tabular}[t]{@{}c@{}}6.599\\(2.606)\end{tabular} & \begin{tabular}[t]{@{}c@{}}0.548\\(0.168)\end{tabular} & \begin{tabular}[t]{@{}c@{}}0.195\\(0.097)\end{tabular} & \begin{tabular}[t]{@{}c@{}}5.408\\(1.359)\end{tabular}\\
Papers & 7,243 & 7,242 & 7,718 & 7,718\\
Acknowledgement length & \begin{tabular}[t]{@{}c@{}}4.968\\(2.573)\end{tabular} & \begin{tabular}[t]{@{}c@{}}0.468\\(0.166)\end{tabular} & \begin{tabular}[t]{@{}c@{}}0.176\\(0.097)\end{tabular} & \begin{tabular}[t]{@{}c@{}}4.478\\(1.372)\end{tabular}\\
Papers & 7,243 & 7,242 & 7,718 & 7,718\\
Full-text baseline & \begin{tabular}[t]{@{}c@{}}6.599\\(2.606)\end{tabular} & \begin{tabular}[t]{@{}c@{}}0.548\\(0.168)\end{tabular} & \begin{tabular}[t]{@{}c@{}}0.240\\(0.093)\end{tabular} & --\\
Papers & 7,243 & 7,242 & 8,502 & --\\
Full-text length & \begin{tabular}[t]{@{}c@{}}5.742\\(2.593)\end{tabular} & \begin{tabular}[t]{@{}c@{}}0.485\\(0.167)\end{tabular} & \begin{tabular}[t]{@{}c@{}}0.218\\(0.094)\end{tabular} & --\\
Papers & 7,243 & 7,242 & 8,502 & --\\
\bottomrule\end{tabular*}}\tabnote{Disclosure coefficients and researcher/paper-clustered standard errors. Each conditioned row is compared with its own identical-sample baseline. Acknowledgement length uses log(1 + words); full-text length uses log(words). Acknowledgement length is partly generated by the credits and disclosures themselves; these are conditional descriptions, not bias-corrected staffing effects. Source and author-dependence limitations apply.}\end{table}

\begin{table}[htbp]\centering\caption{Large lists and document timing}\label{tab:restrictions}
{\small\renewcommand{\arraystretch}{1.12}\setlength{\tabcolsep}{4pt}\begin{tabular*}{\linewidth}{@{\extracolsep{\fill}}lrrrr@{}}
\toprule
 & \shortstack{Any assistant\\(pp)} & \shortstack{Named\\assistants} & Authors & Pages\\\midrule
At most eight assistants & \begin{tabular}[t]{@{}c@{}}6.183\\(2.702)\end{tabular} & \begin{tabular}[t]{@{}c@{}}0.317\\(0.099)\end{tabular} & \begin{tabular}[t]{@{}c@{}}0.213\\(0.106)\end{tabular} & \begin{tabular}[t]{@{}c@{}}4.562\\(1.428)\end{tabular}\\
Papers & 7,131 & 7,131 & 7,131 & 7,131\\
At most ten authors & \begin{tabular}[t]{@{}c@{}}6.658\\(2.631)\end{tabular} & \begin{tabular}[t]{@{}c@{}}0.533\\(0.170)\end{tabular} & \begin{tabular}[t]{@{}c@{}}0.165\\(0.069)\end{tabular} & \begin{tabular}[t]{@{}c@{}}5.476\\(1.275)\end{tabular}\\
Papers & 7,210 & 7,209 & 8,469 & 8,469\\
Independently dated versions & \begin{tabular}[t]{@{}c@{}}-0.840\\(4.258)\end{tabular} & \begin{tabular}[t]{@{}c@{}}0.325\\(0.311)\end{tabular} & \begin{tabular}[t]{@{}c@{}}0.542\\(0.191)\end{tabular} & \begin{tabular}[t]{@{}c@{}}2.173\\(1.869)\end{tabular}\\
Papers & 1,615 & 1,615 & 2,368 & 2,368\\
\bottomrule\end{tabular*}}\tabnote{Disclosure coefficients; standard errors in parentheses cluster by researcher and paper. Eligibility, fixed-effect support and paper-total weights are recomputed after each restriction. The unconditioned main estimates are in Table~\ref{tab:main}. Strict document timing concerns the dated text, not simply a file hash. Smaller samples and dependence limit inference.}\end{table}

\begin{table}[htbp]\centering\caption{Ordinary and equal weights on identical source support}\label{tab:balanced}
{\small\renewcommand{\arraystretch}{1.12}\setlength{\tabcolsep}{4pt}\begin{tabular*}{\linewidth}{@{\extracolsep{\fill}}lrrrr@{}}
\toprule
 & \shortstack{Any assistant\\(pp)} & \shortstack{Named\\assistants} & Authors & Pages\\\midrule
Ordinary paper weights & \begin{tabular}[t]{@{}c@{}}6.544\\(2.665)\end{tabular} & \begin{tabular}[t]{@{}c@{}}0.511\\(0.173)\end{tabular} & \begin{tabular}[t]{@{}c@{}}0.217\\(0.096)\end{tabular} & \begin{tabular}[t]{@{}c@{}}5.146\\(1.323)\end{tabular}\\
Papers & 6,387 & 6,387 & 7,549 & 7,549\\
Equal source weights & \begin{tabular}[t]{@{}c@{}}1.977\\(4.667)\end{tabular} & \begin{tabular}[t]{@{}c@{}}0.361\\(0.168)\end{tabular} & \begin{tabular}[t]{@{}c@{}}0.296\\(0.108)\end{tabular} & \begin{tabular}[t]{@{}c@{}}8.865\\(2.456)\end{tabular}\\
Papers & 6,387 & 6,387 & 7,549 & 7,549\\
\bottomrule\end{tabular*}}\tabnote{Disclosure coefficients and conditional researcher/paper-clustered standard errors. Each outcome retains sources with at least five disclosed and twenty no-statement papers, after renewed fixed-effect pruning. Equal weighting gives each retained source total weight one, divided equally across its papers and then linked-researcher rows. The matched ordinary-weight row isolates reweighting from the support restriction. Source concentration and shared-author dependence remain relevant.}\end{table}

\begin{table}[htbp]\centering\caption{Dependence of disclosure estimates on source composition}\label{tab:omissions}
{\small\renewcommand{\arraystretch}{1.12}\setlength{\tabcolsep}{4pt}\begin{tabular*}{\linewidth}{@{\extracolsep{\fill}}lrrrr@{}}
\toprule
Outcome & Omit NBER & Omit arXiv & \shortstack{Range across\\all omissions} & \shortstack{Sources\\omitted}\\\midrule
Any assistant (pp) & \begin{tabular}[t]{@{}c@{}}3.463\\(3.358)\end{tabular} & \begin{tabular}[t]{@{}c@{}}10.236\\(3.178)\end{tabular} & 3.463 to 10.236 & 24\\
Named assistants & \begin{tabular}[t]{@{}c@{}}0.438\\(0.156)\end{tabular} & \begin{tabular}[t]{@{}c@{}}0.726\\(0.204)\end{tabular} & 0.438 to 0.726 & 24\\
Authors & \begin{tabular}[t]{@{}c@{}}0.276\\(0.163)\end{tabular} & \begin{tabular}[t]{@{}c@{}}0.079\\(0.105)\end{tabular} & 0.079 to 0.338 & 24\\
Pages & \begin{tabular}[t]{@{}c@{}}4.938\\(1.687)\end{tabular} & \begin{tabular}[t]{@{}c@{}}6.005\\(1.601)\end{tabular} & 4.783 to 6.005 & 24\\
\bottomrule\end{tabular*}}\tabnote{Each column refits the within-researcher model after a source omission, with eligibility and paper-total weights recomputed. Parentheses contain conditional researcher/paper-clustered standard errors. Ranges describe point estimates, not confidence bounds. The complete omission estimates, standard errors and sample sizes accompany the results files.}\end{table}

\begin{table}[htbp]\centering\caption{Work-corroborated researcher identities}\label{tab:identity}
{\small\renewcommand{\arraystretch}{1.12}\setlength{\tabcolsep}{4pt}\begin{tabular*}{\linewidth}{@{\extracolsep{\fill}}lrrrr@{}}
\toprule
 & \shortstack{Any assistant\\(pp)} & \shortstack{Named\\assistants} & Authors & Pages\\\midrule
Disclosure & \begin{tabular}[t]{@{}c@{}}7.655\\(2.741)\end{tabular} & \begin{tabular}[t]{@{}c@{}}0.614\\(0.178)\end{tabular} & \begin{tabular}[t]{@{}c@{}}0.201\\(0.097)\end{tabular} & \begin{tabular}[t]{@{}c@{}}5.499\\(1.335)\end{tabular}\\
Papers & 6,953 & 6,952 & 8,135 & 8,135\\
\bottomrule\end{tabular*}}\tabnote{Excludes name-only link extensions and refits the within-researcher comparison. Authorship itself comes from the byline. Parentheses contain conditional researcher/paper-clustered standard errors. Corroboration narrows identity uncertainty but does not establish complete author linkage.}\end{table}

\FloatBarrier

\subsection{Punctuation and historical comparisons}\label{sec:punctuation}
Punctuation offers alternative measurements of text characteristics, with sensitivity to both construction and period. The contemporary family contains 380 distinct coefficient tests across original and cleaned rates, pooled and source-specific historical thresholds, raw-count flags, joint controls, alternative caps, and companion disclosure coefficients. Numerically identical tests are recorded as aliases, not independent evidence. Table~\ref{tab:punctuation} summarises rates under adjustment across that complete family. Continuous coefficients contrast one additional glyph per 1,000 selected words; threshold flags instead compare binary groups. The complete result files preserve those units and all outcomes.

The cleaned-dash coefficient for solo authorship has a full-family adjusted probability of 0.303. It therefore does not establish a smaller-team pattern after accounting for construction choice. Two colon/page comparisons retain adjusted probabilities below 0.001, namely the rate shown in Table~\ref{tab:punctuation} and the same rate in the specification that adds the writing flag and dash rate, but they concern text and document length rather than a verified AI effect. Shared-author and source-dependence qualifications apply to these exploratory comparisons too.

Historical specifications include each text measure and its interaction with the post-2022 period, with researcher, source-quarter, and field effects. Their eighty tests include levels and interactions for four constructions and ten outcomes. The post-2022 interactions supply no adjusted evidence of an AI-specific change. Pre-period associations also demonstrate that text characteristics can relate to research inputs before widespread generative AI. Selection, document timing, and exposition provide competing interpretations; a contemporaneous coefficient does not by itself distinguish them.

\begin{table}[htbp]\centering\caption{Text-rate associations under the complete construction family}\label{tab:punctuation}
{\small\renewcommand{\arraystretch}{1.12}\setlength{\tabcolsep}{4pt}\begin{tabular*}{\linewidth}{@{\extracolsep{\fill}}lrrrr@{}}
\toprule
 & \shortstack{Any assistant\\(pp)} & Authors & Pages & \shortstack{Solo author\\(pp)}\\\midrule
Colon rate & -0.171 & 0.003 & -0.496 & 0.317\\
Full-family Holm & 1.000 & 1.000 & $<0.001$ & 1.000\\
Dash rate & 0.201 & -0.007 & 0.074 & 0.340\\
Full-family Holm & 1.000 & 1.000 & 1.000 & 1.000\\
Cleaned dash rate & 0.201 & -0.022 & 0.182 & 0.649\\
Full-family Holm & 1.000 & 1.000 & 1.000 & 0.303\\
\bottomrule\end{tabular*}}\tabnote{Each rate coefficient contrasts one additional glyph per 1,000 selected words. Joint disclosure controls and the within-researcher design remain included. Holm values cover 380 distinct contemporary coefficient tests, including construction alternatives and companion indicators. Complete pointwise intervals, standard errors and aliases are supplied in the results files. Rates include content/typesetting as well as prose; no AI-specific staffing effect is inferred.}\end{table}

\FloatBarrier

\subsection{Documented attributes}\label{sec:attributes}
An operational disclosure can also establish that a paper uses a computational or data-related method. The full attribute count and the task classification can therefore share evidence. Table~\ref{tab:attributes} removes formal theory, computational execution, and existing-data use, retaining the five acquisition, implementation, and access components. The remaining information is still not guaranteed independent: original data may, for example, include AI annotation. These counts describe documented attributes rather than latent breadth, and neither the full nor reduced comparison validates scope expansion.
\begin{table}[htbp]\centering\caption{Documented attributes and shared evidence}\label{tab:attributes}
{\small\renewcommand{\arraystretch}{1.12}\setlength{\tabcolsep}{4pt}\begin{tabular*}{\linewidth}{@{\extracolsep{\fill}}lrrrr@{}}
\toprule
Index & Indicator & Estimate & SE & Papers\\\midrule
Eight attributes & Writing & -0.083 & 0.064 & 7,251\\
Eight attributes & Disclosure & 0.133 & 0.069 & 7,251\\
Five attributes & Writing & 0.002 & 0.036 & 7,251\\
Five attributes & Disclosure & 0.167 & 0.047 & 7,251\\
\bottomrule\end{tabular*}}\tabnote{The reduced index omits formal theory, computational execution and existing-data use, retaining acquisition, implementation and access attributes. Both indices use identical papers. Standard errors condition on researcher/paper clustering. The reduced comparison is a sensitivity, not an independent measure of scope; original information acquisition may itself use AI-generated annotations. Its pointwise probabilities are not additional confirmatory tests.}\end{table}

\noindent Full estimates, pointwise uncertainty, adjustment families, construction definitions, and sample membership are retained with the replication code and results, including supporting specifications not displayed here.

\subsection{Disclosure wording}\label{sec:wording-appendix}
This section reports the comparison of protective disclosure wording -- explicit limits on AI's intellectual role or claims of human oversight -- summarised in Section 4.2, alongside its sensitivity tables. Across all 1,145 fitted disclosures with source-year and field fixed effects and task and template controls, coauthored papers exceed solo papers in protective wording by 4.9 percentage points (SE 2.0; raw \(p = .022\)). That overall contrast fails the registered eight-test Holm adjustment (\(p = .176\)), which is retained here and in Appendix B.6 because 1.5 to 2.7 effective sources make a source-level bootstrap unreliable. Among the 162 fitted writing-only disclosures, the coauthored-minus-solo difference is 10.2 percentage points (SE 6.3; raw \(p = .143\); wild source-cluster bootstrap \(p = .27\)). ArXiv supplies 80.3\% of the coefficient score mass in that writing-only baseline, and omitting 2026 moves the estimate to \ensuremath{-}0.8 percentage points.

\begin{table}[htbp]\centering\caption{Protective wording in AI disclosures: coauthored minus solo papers}\label{tab:wording}
{\small\renewcommand{\arraystretch}{1.12}\setlength{\tabcolsep}{4pt}\begin{tabular*}{\linewidth}{@{\extracolsep{\fill}}lrrrr@{}}
\toprule
 & \shortstack{Team minus solo\\(pp)} & Papers & \shortstack{Solo\\papers} & Sources\\\midrule
All disclosures & 4.9 (2.0) & 1,145 & -- & 20\\
Writing-only disclosures & 10.2 (6.3) & 162 & 34 & 10\\
\quad Human-oversight component & 11.1 (5.0) & 162 & 34 & 10\\
\quad Limitation component & -7.7 (7.7) & 162 & 34 & 10\\
\quad Named tool or service in the statement & 11.4 (5.0) & 133 & 29 & 8\\
\quad No template-compatible wording & 18.1 (8.7) & 102 & 26 & 9\\
\quad Narrow editing tasks only & 9.7 (7.8) & 106 & 17 & 8\\
\quad Statement-length control & 13.3 (7.0) & 162 & 34 & 10\\
\quad Omitting 2026 & -0.8 (9.7) & 61 & 12 & 6\\
\bottomrule\end{tabular*}}\tabnote{Linear probability models of protective wording among papers with a recorded own-use disclosure. The first row uses all fitted disclosures with source-year and field effects and task and template controls; the remaining rows use writing-only disclosures with source-year and field effects and a template indicator, except where the row removes template-compatible wording. Coefficients are coauthored minus solo in percentage points; source-clustered standard errors in parentheses. The human-oversight and limitation components are the two parts of the composite outcome. A null-restricted wild source-cluster bootstrap over all 1,024 sign patterns gives $p=0.27$ for the writing-only baseline and $p=0.25$ for the narrow-task row. ArXiv supplies 80.3\% of the residual identifying variation in the baseline (1.53 effective sources); the shared-author covariance diagnostic gives a standard error of 9.8 points. Solo counts for the all-disclosure row are not separately recorded after pruning. These are exploratory conditional associations among disclosing papers, not confirmatory tests; Appendix~\ref{sec:wording-appendix} reports every specification.}\end{table}

Protective wording is observed for 1,163 of the 1,167 recorded disclosures; four passages are unresolved. Shares by author count are 50.4\%, 54.4\%, 58.1\%, 49.0\% and 56.6\% for one to five-or-more authors. The registered comparison (T06 in Table~\ref{tab:mechanisms}) fits a paper-level linear probability model on the 1,145 disclosures retained after fixed-effect pruning, with source-year and field effects, task controls, and a template indicator, clustered by source. Its solo-minus-coauthored contrast is $-4.9$ points with an ordinary interval from $-9.0$ to $-0.8$ and a Bonferroni interval from $-11.0$ to $1.1$. Without the task controls the contrast is $-9.3$ points; among operational disclosures alone it is $-3.0$.

The writing-only comparison was a registered secondary split. Under the whole-document task rule, 175 disclosures are writing-only, 119 of them confined to surface editing, proofreading, translation, or typesetting, and all have a known outcome: 103 of 139 coauthored papers (74.1\%) and 25 of 36 solo papers (69.4\%) contain protective language. The estimation sample after singleton pruning has 162 papers from 10 sources: 96 of 128 coauthored (75.0\%) and 23 of 34 solo (67.6\%). Tables~\ref{tab:wording_full} and \ref{tab:wording_full2} report every specification in the follow-up protocol, which was frozen after the secondary estimate was first known and before further coefficients were inspected; the protocol also fixes the restriction to statements naming a tool or service before its estimation. All probabilities are unadjusted for the selection of this secondary comparison. Template-compatible wording means lexical compatibility with the review-and-responsibility formulation in a current publisher policy, not proven copying or a binding rule. The statement-length control uses the log word count of the extracted own-use clauses, which can itself reflect the reporting decision.

Inference is limited by source concentration. ArXiv supplies 80.3\% of the coefficient score mass in the baseline, giving 1.53 effective sources. Every single-source omission retains the sign; excluding arXiv leaves 63 papers with six solo papers and a very wide interval. The restricted wild source-cluster bootstrap enumerates all 1,024 sign patterns and gives $p=0.27$; it is a symmetric-error approximation, not randomisation inference. A shared-author adjacency covariance gives a standard error of 9.8 points against 6.3 under conventional clustering, with 74.7\% of author slots linked. Omitting 2026 leaves 61 papers and a difference of $-0.8$ points. A formal interaction between solo authorship and writing-only task on 1,061 disclosures gives $-0.9$ points (conventional $p=0.88$, wild $p=0.89$); allowing source-year and field effects to differ by task group gives 7.1 points ($p=0.24$, wild $p=0.21$). Raw protective shares among coauthored writing-only papers are 30 of 41, 40 of 50, 17 of 23 and 16 of 25 for two, three, four, and at least five authors, so the evidence concerns the solo-team contrast rather than a gradient.

The reviewing model's first-pass protective label, before source-context adjudication, gives a team-minus-solo difference of 9.4 points on the writing-only sample, compared with 10.2 after adjudication (Table~\ref{tab:wording_full2}). This is a stability check within one measurement procedure, not independent validation. The main author-count sample overlap has only four sources; its point estimate is reported without an interval or probability.

\begin{table}[htbp]\centering\caption{Protective wording among writing-only disclosures: specifications and omissions}\label{tab:wording_full}
{\footnotesize\renewcommand{\arraystretch}{1.08}\setlength{\tabcolsep}{3pt}\begin{tabular*}{\linewidth}{@{\extracolsep{\fill}}lrlrrr@{}}
\toprule
Specification & \shortstack{Team minus\\solo (pp)} & 95\% interval & Papers & Solo & Sources\\\midrule
\multicolumn{6}{@{}l}{\emph{Focused specifications}}\\
Baseline (source-year and field effects, template indicator) & 10.2 (6.3) & [-4.1, 24.4] & 162 & 34 & 10\\
Named tool or service in the own-use statement & 11.4 (5.0) & [-0.3, 23.2] & 133 & 29 & 8\\
Excluding template-compatible wording & 18.1 (8.7) & [-2.0, 38.2] & 102 & 26 & 9\\
Excluding repeated normalised wording & 10.2 (6.3) & [-4.1, 24.4] & 162 & 34 & 10\\
Source-year effects only & 8.1 (6.5) & [-6.6, 22.9] & 163 & 34 & 10\\
Source, year and field effects & 3.3 (7.1) & [-12.4, 19.0] & 170 & 35 & 12\\
Source-year cells with both solo and coauthored papers & 11.3 (6.5) & [-4.0, 26.5] & 124 & 34 & 8\\
Baseline rows without template control & 16.4 (5.3) & [4.3, 28.4] & 162 & 34 & 10\\
Baseline rows with statement-length control & 13.3 (7.0) & [-2.6, 29.1] & 162 & 34 & 10\\
Independently dated documents & 8.2 (n/a) & unavailable & 95 & 27 & 2\\
Unknown outcome coded 0 & 10.2 (6.3) & [-4.1, 24.4] & 162 & 34 & 10\\
Unknown outcome coded 1 & 10.2 (6.3) & [-4.1, 24.4] & 162 & 34 & 10\\
\multicolumn{6}{@{}l}{\emph{Period and source omissions}}\\
Omitting 2024 & 11.2 (6.6) & [-3.7, 26.1] & 148 & 33 & 10\\
Omitting 2025 & 14.7 (8.4) & [-4.7, 34.0] & 110 & 22 & 9\\
Omitting 2026 & -0.8 (9.7) & [-25.7, 24.1] & 61 & 12 & 6\\
Range over the 10 single-source omissions & \multicolumn{2}{l}{7.4 to 30.4} & & & \\
\bottomrule\end{tabular*}}\tabnote{Coefficients are coauthored minus solo in percentage points with conventional source-clustered standard errors and intervals; the sign of the fitted solo coefficient is reversed. The baseline has source-year and field effects and a template indicator. The repeated-wording row applies the frozen normalisation rule. Source-cluster intervals are withheld with fewer than five sources. All probabilities are exploratory and unadjusted for selection of this secondary comparison.}\end{table}

\begin{table}[htbp]\centering\caption{Protective wording among writing-only disclosures: components, interactions, label stability and sample overlap}\label{tab:wording_full2}
{\footnotesize\renewcommand{\arraystretch}{1.08}\setlength{\tabcolsep}{3pt}\begin{tabular*}{\linewidth}{@{\extracolsep{\fill}}lrlrrr@{}}
\toprule
Specification & \shortstack{Team minus\\solo (pp)} & 95\% interval & Papers & Solo & Sources\\\midrule
\multicolumn{6}{@{}l}{\emph{Outcome components (baseline sample)}}\\
Human-oversight component & 11.1 (5.0) & [-0.2, 22.4] & 162 & 34 & 10\\
Limitation component & -7.7 (7.7) & [-25.2, 9.8] & 162 & 34 & 10\\
Responsibility component & -7.3 (6.5) & [-22.1, 7.5] & 162 & 34 & 10\\
\multicolumn{6}{@{}l}{\emph{Task interactions (writing-only and operational/mixed disclosures)}}\\
Writing-only minus operational/mixed, task controls & -0.9 (6.2) & [-14.0, 12.1] & 1,061 & 192 & 20\\
Same, excluding templates & -2.8 (8.1) & [-19.8, 14.3] & 966 & 173 & 20\\
Writing-only minus pure operational & -6.4 (7.3) & [-21.9, 9.1] & 774 & 98 & 18\\
Same, effects varying by task group & 7.1 (5.9) & [-5.2, 19.5] & 1,051 & 190 & 19\\
\multicolumn{6}{@{}l}{\emph{Label stability and narrow editing tasks}}\\
First-pass label before source adjudication, baseline sample & 9.4 (6.1) & [-4.4, 23.3] & 162 & 34 & 10\\
Narrow editing tasks only, protective wording & 9.7 (7.8) & [-8.8, 28.1] & 106 & 17 & 8\\
Narrow editing tasks only, human-oversight component & 5.0 (5.3) & [-7.6, 17.5] & 106 & 17 & 8\\
Narrow editing tasks only, limitation component & 2.7 (10.4) & [-21.9, 27.3] & 106 & 17 & 8\\
\multicolumn{6}{@{}l}{\emph{Overlap with the manuscript author sample}}\\
Papers in the main author-count sample & -19.8 (n/a) & unavailable & 64 & 8 & 4\\
Papers with complete linked rosters & 15.8 (1.3) & [12.1, 19.5] & 98 & 27 & 5\\
Papers with at least one linked author & 12.2 (5.3) & [-1.5, 25.9] & 128 & 28 & 6\\
\bottomrule\end{tabular*}}\tabnote{Same conventions as Table~\ref{tab:wording_full}. Task interactions compare the solo association in writing-only disclosures with operational or mixed-use disclosures; their papers and solo columns count the combined sample. Wild source-cluster probabilities use a null-restricted residual bootstrap with Rademacher signs: 9,999 draws give 0.89 and 0.21 for the first and fourth interaction rows, and enumeration over all sign patterns gives 0.25 for the narrow-task protective row. The first-pass label is the reviewing model's proposal before the source-context adjudication of positive oversight and restriction claims; the comparison is a stability check within one measurement procedure, not independent validation. Sample-overlap rows restrict to papers linked to the manuscript's researcher panel; linking is selective and these are not independent replications.}\end{table}

\subsection{Registered mechanism comparisons}\label{sec:mechanisms-appendix}
\begin{figure}[htbp]\centering
\includegraphics[width=\linewidth]{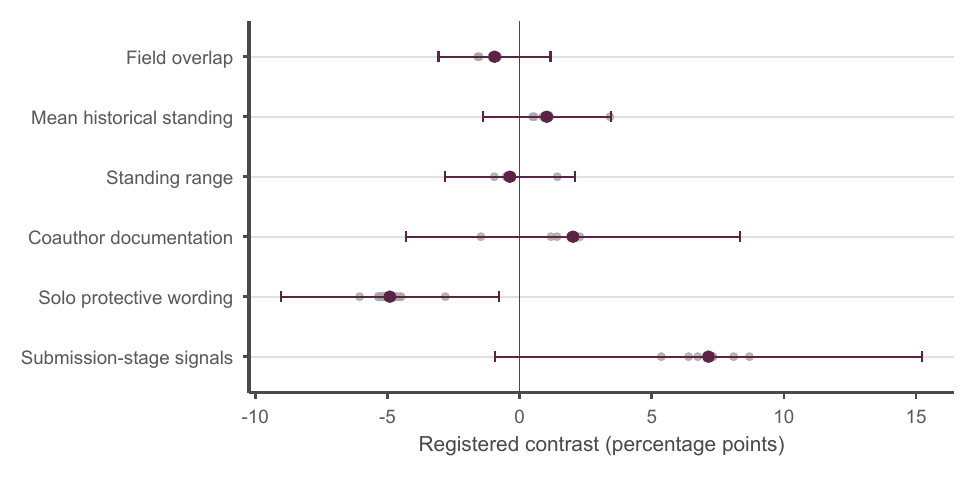}
\caption{Registered contrasts and single-source omissions}\label{fig:mechanisms}
\tabnote{Plum points and bars: registered contrast and ordinary 95\% interval from Table~\ref{tab:mechanisms}. Grey points: the contrast after omitting each source in turn, with eligibility and weights recomputed. T05 without NBER is estimable and is included. Continuous features use the frozen contrast units. No entry survives Holm adjustment, and every entry fails the source-support screen.}
\end{figure}

The register of eight comparisons was frozen before any of their coefficients was estimated, after a feasibility stage that reproduced the main estimates to numerical precision and inventoried coverage. The data had already been analysed, so the freeze is an inductive-analysis protocol, not a preregistration. The registered specifications, contrasts, and family are retained with the adjudicated disclosure labels. All tests except T06 reverse the outcome of the main regressions: they ask which predetermined team characteristics predict a recorded disclosure, in weighted researcher-paper linear probability models with researcher, source-quarter, and field effects, author-count bins, and the prose controls, each paper carrying total weight one. They do not estimate a share of the author-count coefficient explained. Holm adjustment applies across the eight entries, with unestimable entries carried at $p=1$. Table~\ref{tab:mechanisms} and Figure~\ref{fig:mechanisms} report the six estimable entries with ordinary and Bonferroni intervals, adjusted probabilities, every single-source omission, and the concentration of residual information.

\begin{table}[htbp]\centering\caption{Registered comparisons of explanations for disclosure and team size}\label{tab:mechanisms}
{\footnotesize\renewcommand{\arraystretch}{1.1}\setlength{\tabcolsep}{2.5pt}\begin{tabular*}{\linewidth}{@{\extracolsep{\fill}}llrrllrr@{}}
\toprule
Test & Feature & \shortstack{Papers /\\disclosures} & \shortstack{Estimate\\(pp)} & \shortstack{Ordinary\\95\% interval} & \shortstack{Bonferroni\\interval} & \shortstack{Holm\\$p$} & \shortstack{Effective\\sources}\\\midrule
T01 & Field overlap & 2,401 / 87 & -0.95 & [-3.07, 1.17] & [-3.90, 2.01] & 1.000 & 2.51\\
T02 & Mean historical standing & 2,401 / 87 & 1.03 & [-1.40, 3.45] & [-2.36, 4.41] & 1.000 & 2.56\\
T03 & Standing range & 2,401 / 87 & -0.37 & [-2.83, 2.10] & [-3.80, 3.07] & 1.000 & 2.54\\
T05 & Coauthor documentation & 547 / 14 & 2.02 & [-4.30, 8.34] & [-6.82, 10.87] & 1.000 & 1.72\\
T06 & Solo protective wording & 1,145 / 1,145 & -4.91 & [-9.02, -0.79] & [-10.95, 1.14] & 0.176 & 2.05\\
T08 & Submission-stage signals & 8,502 / 544 & 7.15 & [-0.94, 15.23] & [-4.13, 18.43] & 0.582 & 2.72\\
T04 & Prior coauthor propensities & \multicolumn{6}{@{}l}{Unestimable: one paper has complete prior outside-team histories}\\
T07 & Detection recall & \multicolumn{6}{@{}l}{Unidentified: solo minus multi-author recall bounds $[-0.96,0.99]$}\\
\bottomrule\end{tabular*}}\tabnote{Eight entries frozen before estimation and re-estimated unchanged on the reviewed disclosure labels; Holm adjustment across all eight with unestimable entries carried at $p=1$; Bonferroni intervals are simultaneous across the eight. Figure~\ref{fig:mechanisms} shows every single-source omission. All entries except T06 use recorded disclosure as the outcome in weighted researcher-paper linear probability models with researcher, source-quarter and field effects, author-count bins and the prose controls of equation~\ref{eq:empirical}; T01 adds mean standing and career age, T02 and T03 enter mean and range jointly with age. T06 uses protective wording among disclosures with source-year and field effects and task and template controls, clustered by source; other intervals cluster by researcher and paper. Contrasts: T01, pairwise agreement in authors' pre-2022 broad fields, 0 to 1; T02, interquartile change, 24.7 percentile points; T03, interquartile change, 41.6 percentile points; T05, interquartile change in coauthors' pre-2022 log acknowledgement length; T06, solo minus coauthored, among disclosures; T08, presence of venue notice or data and conflict declarations. Source-omission estimates and any explicit failures are retained in the accompanying CSVs. Every estimable entry fails the prespecified source-support screen; effective sources are the inverse sum of squared source shares of residual information.}\end{table}

Historical features require complete linked rosters and observed pre-2022 histories; missing histories are not assigned zeros. Complete rosters cover 794 of 1,167 disclosing and 10,508 of 15,133 other papers, and complete pre-2022 standing covers 205 and 4,701 of them. The field-overlap, standing, and standing-range models therefore describe a selected population of 2,401 papers with 87 disclosures. Standing is a citation percentile within the linked historical population, not a global rank, and field overlap is pairwise agreement in dominant broad fields, a weak proxy for what a reader infers about the operator. Coauthor documentation uses the mean of coauthors' pre-2022 log acknowledgement length; its 14 disclosures leave the estimate dependent on NBER, but omitting NBER remains estimable, with a registered contrast of $-1.46$ percentage points. Prior coauthor disclosure propensities require at least three independently dated earlier papers outside the current team for every coauthor, which one paper satisfies.

The submission-stage indicator is an explicit venue submission or acceptance notice, or the combination of a structured data-availability statement and a conflict-of-interest declaration; the AI passage itself is excluded. Only 98 papers in the 8,502-paper sample carry it, of which 16 disclose, all through paired declarations without an explicit notice; the 17 papers with explicit notices contain no recorded disclosure. On the 547-paper sample with complete coauthor documentation histories, the stage contrast is $1.3$ points before and $1.12$ after adding documentation, with 14 disclosures. On the 7,718-paper sample with acknowledgement text, it is $12.7$ points before and $13.3$ after conditioning on acknowledgement length, and $-6.4$ and $-5.9$ after removing verified AI clauses from the acknowledgements on a 1,485-paper sample. The coefficients also depend on the sample and on removing the AI clauses; pooling the main rows without researcher effects gives $6.1$ points. Secondary comparison outcomes show disclosure associated with $0.49$ more advice or comment credits (interval $-1.62$ to $2.59$) and $-0.2$ points in literal software mentions ($-0.9$ to $0.5$); neither is a guaranteed negative control.

Differential detection cannot be identified from the existing audit. Of its 825 documents, 584 match the current frame and 36 have resolved reference labels. Design-weighted bounds that let unresolved recorded negatives contain missed statements and unresolved positives be false give a solo recall interval of $[0.04,1.00]$ and a multi-author interval of $[0.01,1.00]$, so the difference lies in $[-0.96,0.99]$. A 20-document whole-text pilot covering 1,255 pages left four originally negative documents unresolvable because of garbled mathematics and image-only material, so the planned 300-document text-only audit was not run; a valid reference standard requires page-image adjudication. The second review of candidates described in Section~\ref{sec:measurement-quality} does not supply such a standard, because its candidates were selected by a lexical screen.

The opportunity benchmark is the complementary log-log regression of the disclosure indicator on $\log n$ with source-year cell effects, which the independent constant-probability model restricts to a unit coefficient. The free coefficient is 0.116 (standard error 0.077) on 14,476 papers in 60 cells with outcome variation; 1,824 papers in constant cells are excluded, and overlapping authors are not fully addressed by its covariance. The conditional author-count curve in Figure~\ref{fig:teamsize} uses the 8,502-paper sample; its all-author slope of 1.36 points per doubling has a source-clustered interval from $-0.67$ to $3.38$, and the coauthored-only slope of 2.26 points on a re-pruned 7,599-paper sample has a source-clustered interval from $-1.08$ to $5.60$. Descriptive standardisation on common source-year-field support leaves solo and pair rates of 8.9\% and 7.3\%; adding historical career-age support gives 6.6\% and 3.8\% on a smaller population. Among solo papers with assessable credits, disclosure is 10.5\% without a named assistant credit and 6.1\% with one.

\subsection{Assistant-credit portfolios}\label{sec:portfolios}
Credited assistance can be reused across a researcher's projects, so credits per paper need not track distinct people. For each linked researcher and calendar year, the portfolio comprises their papers $Q$ in the 22 sources represented in every year from 2015 to 2025, the credited assistant-paper links $L$, and the distinct assistants $H$, with credits per paper $a=L/Q$ and papers per assistant $r=L/H$, so that $H=Qa/r$. Complete acknowledgement coverage is required; 98,577 of 107,836 candidate portfolios are retained across the full-year and January--September windows, which overlap. Cross-paper assistant identities lack corroborating institution and date evidence, so $H$ is reported as identity-free bounds between the largest single-paper count and the total number of credits, with exact-name merging shown as an unverified scenario. These portfolios do not depend on disclosure measurement.

\begin{figure}[htbp]\centering
\includegraphics[width=\linewidth]{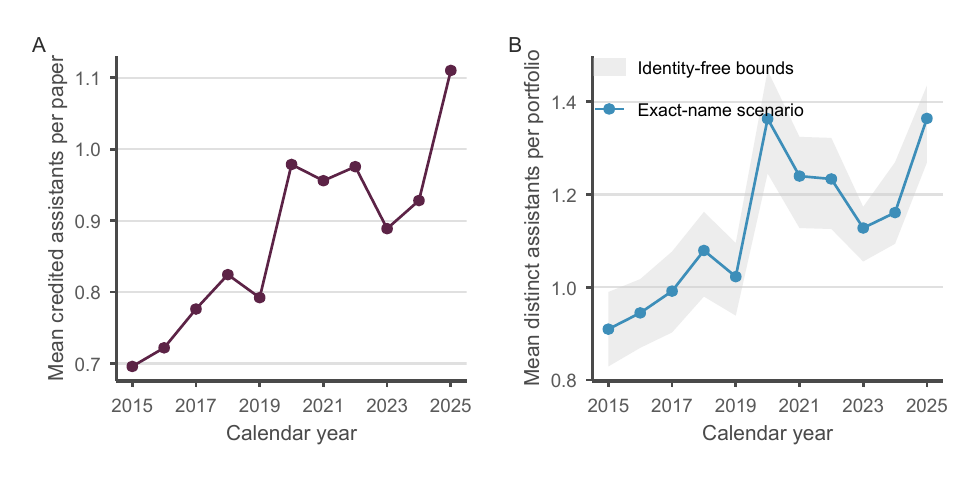}
\caption{Credited assistants within annual researcher portfolios}\label{fig:portfolios}
\tabnote{Full-year portfolios, 2015--2025, of linked researchers with complete acknowledgement coverage in sources present in every year. Panel A: mean credited assistant-paper links per paper. Panel B: mean distinct assistants per portfolio under the unverified exact-name scenario, with identity-free bounds in grey. Researcher composition changes across years, portfolios of coauthors overlap and cannot be summed as a workforce, and counts refer to credits rather than employment or hours.}
\end{figure}

Mean credits per paper rise from 0.98 in 2022 to 1.11 in 2025, with a dip to 0.89 in 2023. For 2025 the identity-free bounds on distinct assistants per portfolio are 1.27 to 1.43. The January--September 2026 window is compared only with the same months of earlier years. These descriptions do not establish increased reuse, employment displacement, or an AI effect; the cross-paper reuse rate remains unresolved until assistant identities are corroborated.

\FloatBarrier
\subsection{Exploratory decompositions and step-down inference}\label{sec:exploratory}
\begin{figure}[htbp]\centering
\includegraphics[width=\linewidth]{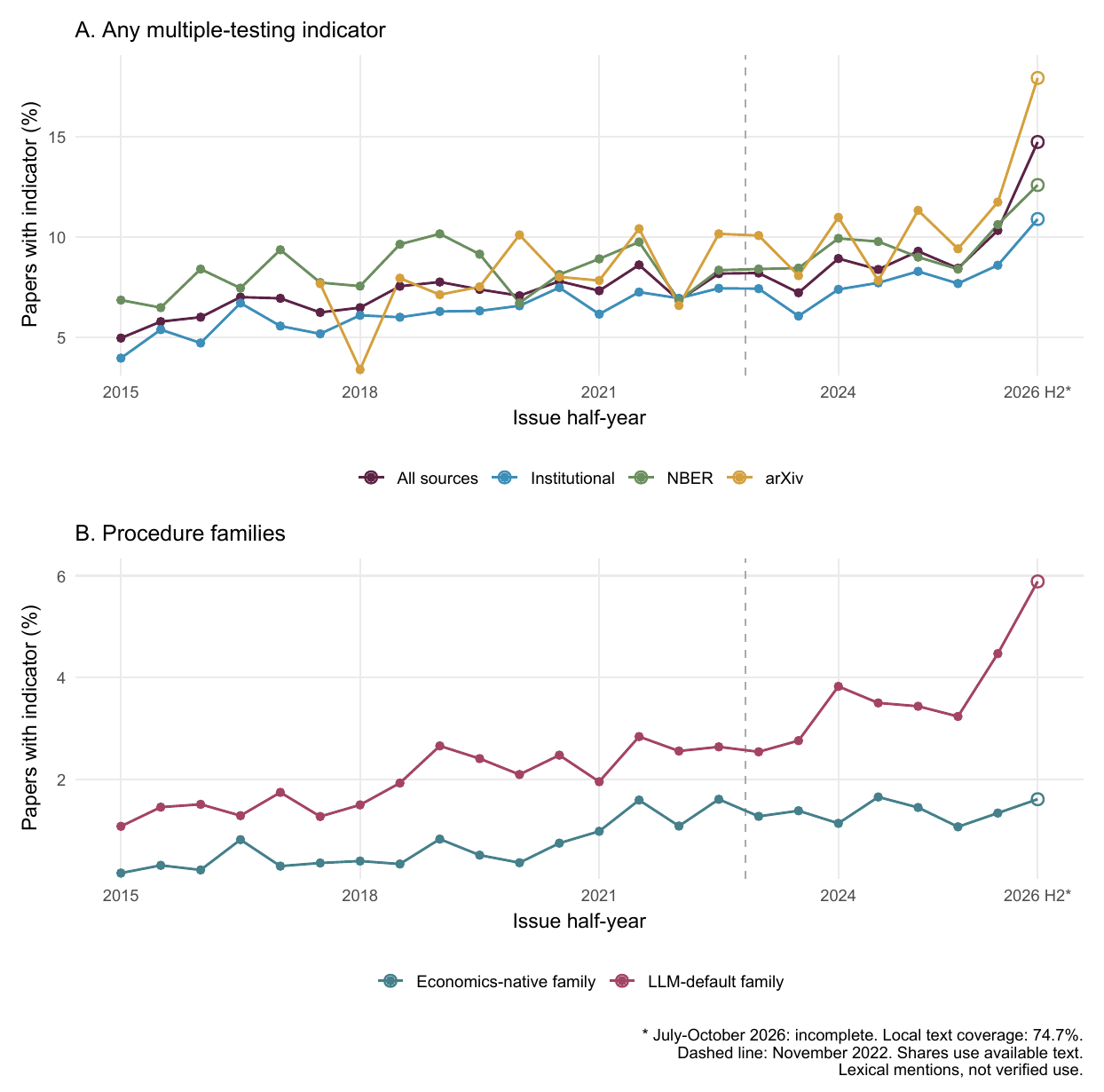}
\caption{Mentions of multiple-testing corrections}\label{fig:mht}
\tabnote{Half-year shares of timing-supported papers with available text whose text mentions a multiple-testing correction outside the reference list. Panel A: any correction by source group. Panel B: the LLM-default family (Holm, Bonferroni, Benjamini--Hochberg) against the economics-native family (Romano--Wolf, List--Shaikh--Xu, Westfall--Young, Anderson). The dashed line marks November 2022. The last cohort is incomplete, with 74.7\% text coverage. These are lexical mentions, not verified use. These exploratory comparisons sit outside the registered twenty-test family; reported p-values are unadjusted.}\end{figure}

This section collects the supporting tables and figure for the analyses added after the registered protocol was frozen, reporting raw \(p\)-values for every exploratory outcome and indicator split. Task-mix regressions appear first, pairing descriptive means with pooled and within-researcher estimates for operational-or-mixed and writing-only disclosures outside the registered twenty-test family. Page-split diagnostics follow across 8,493 verified texts, reporting 60-paper boundary-audit error rates alongside single-source omissions and estimates on the 1,225 bookmark-confirmed papers. Multiple-testing results and Figure~\ref{fig:mht} cover 41,828 papers, pairing 160-snippet precision checks and half-year shares since 2015 with post-2022 source-year and 8,193-paper within-researcher models. Last, the step-down inference table sets 1,999-draw researcher-bootstrap Romano--Wolf \(p\)-values beside the registered Holm values across all twenty main tests and the four reference-list tests.

\begin{table}[htbp]\centering\caption{Task mix: pooled comparisons}\label{tab:taskmix-pooled}
{\small\renewcommand{\arraystretch}{1.12}\setlength{\tabcolsep}{4pt}\begin{tabular*}{\linewidth}{@{\extracolsep{\fill}}lrrr@{}}\toprule
Outcome & Operational & Writing-only & Equality $p$\\\midrule
Any assistant (pp) & \begin{tabular}[t]{@{}c@{}}11.149 (2.591)\\$<.0001$\end{tabular} & \begin{tabular}[t]{@{}c@{}}6.118 (4.959)\\.2173\end{tabular} & .3558\\
Named assistants & \begin{tabular}[t]{@{}c@{}}0.753 (0.173)\\$<.0001$\end{tabular} & \begin{tabular}[t]{@{}c@{}}0.160 (0.214)\\.4539\end{tabular} & .0233\\
Authors & \begin{tabular}[t]{@{}c@{}}0.314 (0.104)\\.0026\end{tabular} & \begin{tabular}[t]{@{}c@{}}0.345 (0.177)\\.051\end{tabular} & .8747\\
Total pages & \begin{tabular}[t]{@{}c@{}}7.113 (1.384)\\$<.0001$\end{tabular} & \begin{tabular}[t]{@{}c@{}}0.571 (2.225)\\.7975\end{tabular} & .0108\\
\bottomrule\end{tabular*}}\tabnote{Pooled regressions omit researcher effects but retain source-quarter and field effects, the writing flag and prose controls. Entries show coefficients, researcher/paper-clustered standard errors and raw p-values; the last column tests equality. These exploratory comparisons sit outside the registered twenty-test family; reported p-values are unadjusted.}\end{table}

\FloatBarrier
\begin{table}[htbp]\centering\caption{Task mix: descriptive means}\label{tab:taskmix-means}
{\small\renewcommand{\arraystretch}{1.12}\setlength{\tabcolsep}{4pt}\begin{tabular*}{\linewidth}{@{\extracolsep{\fill}}lrrr@{}}\toprule
Outcome & Disclosure group & Papers & Mean\\\midrule
Any assistant (pp) & no-disclosure & 6,775 & 28.753\\
Any assistant (pp) & operational-or-mixed & 361 & 37.119\\
Any assistant (pp) & writing-only & 68 & 23.529\\
Any assistant (pp) & unlabelled & 39 & 15.385\\
Named assistants & no-disclosure & 6,774 & 0.829\\
Named assistants & operational-or-mixed & 361 & 1.518\\
Named assistants & writing-only & 68 & 0.647\\
Named assistants & unlabelled & 39 & 0.385\\
Authors & no-disclosure & 7,958 & 3.078\\
Authors & operational-or-mixed & 425 & 3.327\\
Authors & writing-only & 78 & 3.295\\
Authors & unlabelled & 41 & 1.902\\
Total pages & no-disclosure & 7,958 & 52.430\\
Total pages & operational-or-mixed & 425 & 62.259\\
Total pages & writing-only & 78 & 49.654\\
Total pages & unlabelled & 41 & 43.439\\
\bottomrule\end{tabular*}}\tabnote{Paper-level, unadjusted means by disclosure group. Assistant incidence uses percentage points; other outcomes use counts or pages. These exploratory comparisons sit outside the registered twenty-test family; reported p-values are unadjusted.}\end{table}

\FloatBarrier
\begin{table}[htbp]\centering\caption{Task mix: support and within-researcher variation}\label{tab:taskmix-support}
{\small\renewcommand{\arraystretch}{1.12}\setlength{\tabcolsep}{4pt}\begin{tabular*}{\linewidth}{@{\extracolsep{\fill}}lrrrrrr@{}}\toprule
Outcome & O & W & U & \shortstack{O\\switchers} & \shortstack{W\\switchers} & Both\\\midrule
Any assistant (pp) & 361 & 68 & 39 & 441 & 90 & 13\\
Named assistants & 361 & 68 & 39 & 441 & 90 & 13\\
Authors & 425 & 78 & 41 & 521 & 111 & 17\\
Total pages & 425 & 78 & 41 & 521 & 111 & 17\\
\bottomrule\end{tabular*}}\tabnote{O, W and U count operational-or-mixed, writing-only and unlabelled disclosed papers. Switchers are researchers with within-portfolio variation in the indicated dummy. Both counts researchers observing both O and W papers. These exploratory comparisons sit outside the registered twenty-test family; reported p-values are unadjusted.}\end{table}

\FloatBarrier
\begin{table}[htbp]\centering\caption{Page boundaries: restricted-sample disclosure estimates}\label{tab:pagesplit-robustness}
{\small\renewcommand{\arraystretch}{1.12}\setlength{\tabcolsep}{4pt}\begin{tabular*}{\linewidth}{@{\extracolsep{\fill}}p{1.6in}lrrr@{}}\toprule
Sample & Outcome & Estimate (SE), $p$ & Papers & Disclosed\\\midrule
Omit NBER & Main-text pages & \begin{tabular}[t]{@{}c@{}}0.26 (0.92)\\.7752\end{tabular} & 5,131 & 343\\
Omit NBER & Appendix pages & \begin{tabular}[t]{@{}c@{}}4.53 (1.47)\\.002\end{tabular} & 5,131 & 343\\
Omit arXiv & Main-text pages & \begin{tabular}[t]{@{}c@{}}2.48 (0.83)\\.0027\end{tabular} & 6,553 & 341\\
Omit arXiv & Appendix pages & \begin{tabular}[t]{@{}c@{}}3.41 (1.40)\\.0149\end{tabular} & 6,553 & 341\\
Bookmark-confirmed & Main-text pages & \begin{tabular}[t]{@{}c@{}}4.07 (1.89)\\.0313\end{tabular} & 1,225 & 48\\
Bookmark-confirmed & Appendix pages & \begin{tabular}[t]{@{}c@{}}6.23 (3.57)\\.0813\end{tabular} & 1,225 & 48\\
\bottomrule\end{tabular*}}\tabnote{Frozen pages-sample weights, original controls and fixed effects; researcher/paper-clustered standard errors. Raw p-values accompany each coefficient. These exploratory comparisons sit outside the registered twenty-test family; reported p-values are unadjusted.}\end{table}

\FloatBarrier
\begin{table}[htbp]\centering\caption{Held-out page-boundary validation}\label{tab:pagesplit-validation}
{\small\renewcommand{\arraystretch}{1.12}\setlength{\tabcolsep}{4pt}\begin{tabular*}{\linewidth}{@{\extracolsep{\fill}}lrrr@{}}\toprule
Boundary & Reviewed & Errors & Error rate (\%)\\\midrule
Appendix boundary & 60 & 7 & 11.7\\
References boundary & 60 & 4 & 6.7\\
\bottomrule\end{tabular*}}\tabnote{Held-out documentary checks on papers not used for tuning. Unweighted rates; weighting by prevalence gives an appendix-boundary error rate of 13.3\%. These exploratory comparisons sit outside the registered twenty-test family; reported p-values are unadjusted.}\end{table}

\FloatBarrier
\FloatBarrier
\begingroup\footnotesize\setlength{\parindent}{0pt}\setlength{\tabcolsep}{4pt}\setlength{\LTleft}{0pt}\begin{longtable}{lrrrrr}
\caption{Correction mentions by half-year: all sources}\label{tab:mht-trends}\\\toprule
Half-year & Text papers & \shortstack{Any correction\\(\%)} & Holm (\%) & \shortstack{LLM-default\\family (\%)} & \shortstack{Economics-native\\family (\%)}\\\midrule
\endfirsthead\toprule
Half-year & Text papers & \shortstack{Any correction\\(\%)} & Holm (\%) & \shortstack{LLM-default\\family (\%)} & \shortstack{Economics-native\\family (\%)}\\\midrule
\endhead
2015 H1 & 1,305 & 4.98 & 0.00 & 1.07 & 0.15\\
2015 H2 & 1,310 & 5.80 & 0.23 & 1.45 & 0.31\\
2016 H1 & 1,395 & 6.02 & 0.07 & 1.51 & 0.22\\
2016 H2 & 1,481 & 7.02 & 0.20 & 1.28 & 0.81\\
2017 H1 & 1,379 & 6.96 & 0.15 & 1.74 & 0.29\\
2017 H2 & 1,422 & 6.26 & 0.35 & 1.27 & 0.35\\
2018 H1 & 1,539 & 6.50 & 0.06 & 1.49 & 0.39\\
2018 H2 & 1,507 & 7.56 & 0.40 & 1.92 & 0.33\\
2019 H1 & 1,582 & 7.77 & 0.51 & 2.65 & 0.82\\
2019 H2 & 1,580 & 7.41 & 0.32 & 2.41 & 0.51\\
2020 H1 & 1,960 & 7.09 & 0.56 & 2.09 & 0.36\\
2020 H2 & 2,022 & 7.81 & 0.30 & 2.47 & 0.74\\
2021 H1 & 1,948 & 7.34 & 0.21 & 1.95 & 0.98\\
2021 H2 & 1,762 & 8.63 & 0.40 & 2.84 & 1.59\\
2022 H1 & 1,762 & 6.87 & 0.34 & 2.55 & 1.08\\
2022 H2 & 1,745 & 8.19 & 0.17 & 2.64 & 1.60\\
2023 H1 & 1,812 & 8.22 & 0.11 & 2.54 & 1.27\\
2023 H2 & 1,740 & 7.24 & 0.23 & 2.76 & 1.38\\
2024 H1 & 1,856 & 8.94 & 0.75 & 3.83 & 1.13\\
2024 H2 & 2,000 & 8.40 & 0.35 & 3.50 & 1.65\\
2025 H1 & 2,009 & 9.31 & 0.35 & 3.43 & 1.44\\
2025 H2 & 2,257 & 8.46 & 0.31 & 3.23 & 1.06\\
2026 H1 & 2,774 & 10.35 & 0.61 & 4.47 & 1.33\\
2026 H2 & 1,681 & 14.75 & 1.55 & 5.89 & 1.61\\
\bottomrule\end{longtable}\tabnote{Timing-supported papers with available text; mentions outside reference lists, not verified use. LLM-default family (Holm, Bonferroni, Benjamini--Hochberg); Economics-native family (Romano--Wolf, List--Shaikh--Xu, Westfall--Young, Anderson). The final half-year is incomplete. These exploratory comparisons sit outside the registered twenty-test family; reported p-values are unadjusted.}\par\endgroup

\FloatBarrier
\begin{table}[htbp]\centering\caption{Correction mentions and recorded AI indicators}\label{tab:mht-overlap}
{\small\renewcommand{\arraystretch}{1.12}\setlength{\tabcolsep}{4pt}\begin{tabular*}{\linewidth}{@{\extracolsep{\fill}}>{\raggedright\arraybackslash}p{1.7in}rrrrr@{}}\toprule
Group & Papers & \shortstack{Any correction\\(count)} & \shortstack{Any correction\\(\%)} & \shortstack{Holm\\(count)} & Holm (\%)\\\midrule
All observed & 13,225 & 1,291 & 9.76 & 67 & 0.51\\
Disclosed & 891 & 123 & 13.80 & 12 & 1.35\\
Writing flagged & 1,103 & 110 & 9.97 & 8 & 0.73\\
Operational or mixed disclosure & 693 & 96 & 13.85 & 10 & 1.44\\
Writing-only disclosure & 133 & 17 & 12.78 & 2 & 1.50\\
Unclear disclosure & 65 & 10 & 15.38 & 0 & 0.00\\
\bottomrule\end{tabular*}}\tabnote{Post-2022 papers jointly observed on the indicators; groups overlap. Counts and shares refer to lexical mentions. These exploratory comparisons sit outside the registered twenty-test family; reported p-values are unadjusted.}\end{table}

\FloatBarrier
\begin{table}[htbp]\centering\caption{Correction mentions: source year estimates (1)}\label{tab:mht-model-1}
{\small\renewcommand{\arraystretch}{1.12}\setlength{\tabcolsep}{4pt}\begin{tabular*}{\linewidth}{@{\extracolsep{\fill}}>{\raggedright\arraybackslash}p{.95in}>{\raggedright\arraybackslash}p{1.65in}rrrr@{}}\toprule
Outcome & Regressor & \shortstack{Coef.\\(pp)} & \shortstack{SE\\(pp)} & Raw $p$ & $N$\\\midrule
Holm & Operational disclosure & 0.742 & 0.340 & .032 & 13,225\\
Holm & Writing-only disclosure & 0.895 & 0.706 & .209 & 13,225\\
Holm & Unclear disclosure & -0.569 & 0.106 & $< .001$ & 13,225\\
Holm & Writing flag & 0.168 & 0.251 & .505 & 13,225\\
Any correction & Operational disclosure & 2.160 & 1.116 & .056 & 13,225\\
Any correction & Writing-only disclosure & 1.459 & 1.636 & .375 & 13,225\\
Any correction & Unclear disclosure & 3.021 & 5.557 & .588 & 13,225\\
Any correction & Writing flag & -0.281 & 1.126 & .804 & 13,225\\
\bottomrule\end{tabular*}}\tabnote{The indicators enter jointly; operational and writing-only estimates control separately for unlabelled disclosures. Coefficients and standard errors are percentage points. Prose controls omitted. Source-year and field effects; N counts papers. Standard errors cluster by source-year. These exploratory comparisons sit outside the registered twenty-test family; reported p-values are unadjusted.}\end{table}

\FloatBarrier
\begin{table}[htbp]\centering\caption{Correction mentions: source year estimates (2)}\label{tab:mht-model-2}
{\small\renewcommand{\arraystretch}{1.12}\setlength{\tabcolsep}{4pt}\begin{tabular*}{\linewidth}{@{\extracolsep{\fill}}>{\raggedright\arraybackslash}p{.95in}>{\raggedright\arraybackslash}p{1.65in}rrrr@{}}\toprule
Outcome & Regressor & \shortstack{Coef.\\(pp)} & \shortstack{SE\\(pp)} & Raw $p$ & $N$\\\midrule
Holm & Operational disclosure & 0.742 & 0.256 & .008 & 13,225\\
Holm & Writing-only disclosure & 0.895 & 0.500 & .087 & 13,225\\
Holm & Unclear disclosure & -0.569 & 0.083 & $< .001$ & 13,225\\
Holm & Writing flag & 0.168 & 0.327 & .612 & 13,225\\
Any correction & Operational disclosure & 2.160 & 1.103 & .062 & 13,225\\
Any correction & Writing-only disclosure & 1.459 & 1.969 & .466 & 13,225\\
Any correction & Unclear disclosure & 3.021 & 4.630 & .520 & 13,225\\
Any correction & Writing flag & -0.281 & 1.275 & .828 & 13,225\\
\bottomrule\end{tabular*}}\tabnote{The indicators enter jointly; operational and writing-only estimates control separately for unlabelled disclosures. Coefficients and standard errors are percentage points. Prose controls omitted. Source-year and field effects; N counts papers. Standard errors cluster by source. These exploratory comparisons sit outside the registered twenty-test family; reported p-values are unadjusted.}\end{table}

\FloatBarrier
\begin{table}[htbp]\centering\caption{Correction mentions: source year estimates (3)}\label{tab:mht-model-3}
{\small\renewcommand{\arraystretch}{1.12}\setlength{\tabcolsep}{4pt}\begin{tabular*}{\linewidth}{@{\extracolsep{\fill}}>{\raggedright\arraybackslash}p{.95in}>{\raggedright\arraybackslash}p{1.65in}rrrr@{}}\toprule
Outcome & Regressor & \shortstack{Coef.\\(pp)} & \shortstack{SE\\(pp)} & Raw $p$ & $N$\\\midrule
Holm & Operational disclosure & 0.722 & 0.343 & .038 & 13,225\\
Holm & Writing-only disclosure & 0.933 & 0.698 & .185 & 13,225\\
Holm & Unclear disclosure & -0.555 & 0.128 & $< .001$ & 13,225\\
Holm & Writing flag & 0.164 & 0.254 & .519 & 13,225\\
Holm & Log words & -0.527 & 0.680 & .440 & 13,225\\
Holm & Log windows & 0.980 & 0.605 & .109 & 13,225\\
Any correction & Operational disclosure & 1.883 & 1.050 & .076 & 13,225\\
Any correction & Writing-only disclosure & 1.979 & 1.674 & .240 & 13,225\\
Any correction & Unclear disclosure & 3.374 & 5.454 & .538 & 13,225\\
Any correction & Writing flag & -0.275 & 1.106 & .804 & 13,225\\
Any correction & Log words & -0.550 & 2.661 & .837 & 13,225\\
Any correction & Log windows & 6.808 & 2.481 & .007 & 13,225\\
\bottomrule\end{tabular*}}\tabnote{The indicators enter jointly; operational and writing-only estimates control separately for unlabelled disclosures. Coefficients and standard errors are percentage points. Prose controls included. Source-year and field effects; N counts papers. Standard errors cluster by source-year. These exploratory comparisons sit outside the registered twenty-test family; reported p-values are unadjusted.}\end{table}

\FloatBarrier
\begin{table}[htbp]\centering\caption{Correction mentions: source year estimates (4)}\label{tab:mht-model-4}
{\small\renewcommand{\arraystretch}{1.12}\setlength{\tabcolsep}{4pt}\begin{tabular*}{\linewidth}{@{\extracolsep{\fill}}>{\raggedright\arraybackslash}p{.95in}>{\raggedright\arraybackslash}p{1.65in}rrrr@{}}\toprule
Outcome & Regressor & \shortstack{Coef.\\(pp)} & \shortstack{SE\\(pp)} & Raw $p$ & $N$\\\midrule
Holm & Operational disclosure & 0.722 & 0.262 & .011 & 13,225\\
Holm & Writing-only disclosure & 0.933 & 0.502 & .076 & 13,225\\
Holm & Unclear disclosure & -0.555 & 0.096 & $< .001$ & 13,225\\
Holm & Writing flag & 0.164 & 0.326 & .620 & 13,225\\
Holm & Log words & -0.527 & 0.579 & .372 & 13,225\\
Holm & Log windows & 0.980 & 0.524 & .074 & 13,225\\
Any correction & Operational disclosure & 1.883 & 0.979 & .067 & 13,225\\
Any correction & Writing-only disclosure & 1.979 & 1.938 & .318 & 13,225\\
Any correction & Unclear disclosure & 3.374 & 4.429 & .454 & 13,225\\
Any correction & Writing flag & -0.275 & 1.213 & .822 & 13,225\\
Any correction & Log words & -0.550 & 3.946 & .890 & 13,225\\
Any correction & Log windows & 6.808 & 3.723 & .080 & 13,225\\
\bottomrule\end{tabular*}}\tabnote{The indicators enter jointly; operational and writing-only estimates control separately for unlabelled disclosures. Coefficients and standard errors are percentage points. Prose controls included. Source-year and field effects; N counts papers. Standard errors cluster by source. These exploratory comparisons sit outside the registered twenty-test family; reported p-values are unadjusted.}\end{table}

\FloatBarrier
\begin{table}[htbp]\centering\caption{Correction mentions: within researcher estimates (5)}\label{tab:mht-model-5}
{\small\renewcommand{\arraystretch}{1.12}\setlength{\tabcolsep}{4pt}\begin{tabular*}{\linewidth}{@{\extracolsep{\fill}}>{\raggedright\arraybackslash}p{.95in}>{\raggedright\arraybackslash}p{1.65in}rrrr@{}}\toprule
Outcome & Regressor & \shortstack{Coef.\\(pp)} & \shortstack{SE\\(pp)} & Raw $p$ & $N$\\\midrule
Holm & Writing flag & 1.098 & 0.696 & .115 & 15,421\\
Holm & Disclosed & -0.287 & 0.638 & .653 & 15,421\\
Holm & Log words & -0.015 & 0.920 & .987 & 15,421\\
Holm & Log windows & 0.763 & 0.813 & .348 & 15,421\\
Any correction & Writing flag & 2.374 & 1.834 & .196 & 15,421\\
Any correction & Disclosed & 0.650 & 1.936 & .737 & 15,421\\
Any correction & Log words & -1.356 & 3.544 & .702 & 15,421\\
Any correction & Log windows & 6.616 & 3.400 & .052 & 15,421\\
\bottomrule\end{tabular*}}\tabnote{Writing and disclosure indicators enter jointly. Coefficients and standard errors are percentage points. Researcher and paper clusters, the original pages-sample expansion and paper-total weights. N counts researcher-paper rows, covering 8,193 distinct papers. These exploratory comparisons sit outside the registered twenty-test family; reported p-values are unadjusted.}\end{table}

\FloatBarrier
\begin{table}[htbp]\centering\caption{Correction mentions: within researcher estimates (6)}\label{tab:mht-model-6}
{\small\renewcommand{\arraystretch}{1.12}\setlength{\tabcolsep}{4pt}\begin{tabular*}{\linewidth}{@{\extracolsep{\fill}}>{\raggedright\arraybackslash}p{.95in}>{\raggedright\arraybackslash}p{1.65in}rrrr@{}}\toprule
Outcome & Regressor & \shortstack{Coef.\\(pp)} & \shortstack{SE\\(pp)} & Raw $p$ & $N$\\\midrule
Holm & Writing flag & 1.100 & 0.699 & .116 & 15,421\\
Holm & Operational disclosure & -0.417 & 0.616 & .498 & 15,421\\
Holm & Writing-only disclosure & 0.569 & 2.173 & .793 & 15,421\\
Holm & Unclear disclosure & -0.810 & 0.554 & .144 & 15,421\\
Holm & Log words & -0.020 & 0.923 & .983 & 15,421\\
Holm & Log windows & 0.769 & 0.815 & .345 & 15,421\\
Any correction & Writing flag & 2.304 & 1.793 & .199 & 15,421\\
Any correction & Operational disclosure & 1.349 & 2.152 & .531 & 15,421\\
Any correction & Writing-only disclosure & -5.747 & 4.380 & .190 & 15,421\\
Any correction & Unclear disclosure & 8.677 & 10.918 & .427 & 15,421\\
Any correction & Log words & -1.263 & 3.533 & .721 & 15,421\\
Any correction & Log windows & 6.519 & 3.387 & .054 & 15,421\\
\bottomrule\end{tabular*}}\tabnote{The indicators enter jointly; operational and writing-only estimates control separately for unlabelled disclosures. Coefficients and standard errors are percentage points. Researcher and paper clusters, the original pages-sample expansion and paper-total weights. N counts researcher-paper rows, covering 8,193 distinct papers. These exploratory comparisons sit outside the registered twenty-test family; reported p-values are unadjusted.}\end{table}

\FloatBarrier
\begin{table}[htbp]\centering\caption{Correction mentions relative to the pre-2022 trend}\label{tab:mht-pretrend}
{\small\renewcommand{\arraystretch}{1.12}\setlength{\tabcolsep}{4pt}\begin{tabular*}{\linewidth}{@{\extracolsep{\fill}}>{\raggedright\arraybackslash}p{2.55in}rrrr@{}}\toprule
Indicator & Papers & Observed (\%) & Trend (\%) & Gap (pp)\\\midrule
Any correction & 12,577 & 9.91 & 9.06 & 0.85\\
Holm & 12,577 & 0.62 & 0.47 & 0.15\\
LLM-default family (Holm, Bonferroni, Benjamini--Hochberg) & 12,577 & 4.02 & 3.33 & 0.70\\
Economics-native family (Romano--Wolf, List--Shaikh--Xu, Westfall--Young, Anderson) & 12,577 & 1.36 & 1.67 & -0.31\\
\bottomrule\end{tabular*}}\tabnote{Linear half-year trends fitted over 2015 H1--2022 H2 and extrapolated to the pooled 2024--2026 sample. Gaps are descriptive deviations, not break tests. These exploratory comparisons sit outside the registered twenty-test family; reported p-values are unadjusted.}\end{table}

\FloatBarrier
\begin{table}[htbp]\centering\caption{Hit-level precision of correction mentions}\label{tab:mht-precision}
{\small\renewcommand{\arraystretch}{1.12}\setlength{\tabcolsep}{4pt}\begin{tabular*}{\linewidth}{@{\extracolsep{\fill}}>{\raggedright\arraybackslash}p{2.55in}rrrr@{}}\toprule
Indicator & Reviewed & Valid & Precision (\%) & Explicit use\\\midrule
Anderson sharpened q-values & 7 & 7 & 100.0 & 7\\
Benjamini--Hochberg or FDR & 20 & 17 & 85.0 & 15\\
Bonferroni & 20 & 18 & 90.0 & 16\\
Generic phrases & 40 & 24 & 60.0 & 19\\
Holm & 50 & 49 & 98.0 & 44\\
List--Shaikh--Xu & 8 & 8 & 100.0 & 6\\
Romano--Wolf & 8 & 7 & 87.5 & 7\\
Westfall--Young & 7 & 7 & 100.0 & 7\\
\bottomrule\end{tabular*}}\tabnote{Audited snippets, grouped by lexical indicator. Valid correction mentions need not establish actual application; the final column separately counts explicit use. Hit-level precision does not measure recall. These exploratory comparisons sit outside the registered twenty-test family; reported p-values are unadjusted.}\end{table}

\FloatBarrier
\begin{table}[htbp]\centering\caption{Step-down inference for main and reference outcomes}\label{tab:rw}
{\small\renewcommand{\arraystretch}{1.12}\setlength{\tabcolsep}{4pt}\begin{tabular*}{\linewidth}{@{\extracolsep{\fill}}lp{1.65in}lrrr@{}}\toprule
Family & Outcome & Indicator & Raw $p$ & RW $p$ & Holm $p$\\\midrule
Main & Any assistant (pp) & Writing & .155 & .819 & 1.000\\
Main & Any assistant (pp) & Disclosed & .011 & .157 & .170\\
Main & Named assistants & Writing & .236 & .898 & 1.000\\
Main & Named assistants & Disclosed & .001 & .020 & .019\\
Main & Authors & Writing & .863 & .998 & 1.000\\
Main & Authors & Disclosed & .010 & .147 & .163\\
Main & Total pages & Writing & .529 & .994 & 1.000\\
Main & Total pages & Disclosed & $< .001$ & $< .001$ & $< .001$\\
Main & Documented attributes & Writing & .193 & .862 & 1.000\\
Main & Documented attributes & Disclosed & .055 & .488 & .772\\
Main & Solo author & Writing & .299 & .921 & 1.000\\
Main & Solo author & Disclosed & .829 & .998 & 1.000\\
Main & Solo, no assistance & Writing & .432 & .979 & 1.000\\
Main & Solo, no assistance & Disclosed & .917 & .998 & 1.000\\
Main & Any reader & Writing & .759 & .998 & 1.000\\
Main & Any reader & Disclosed & .648 & .994 & 1.000\\
Main & Log named assistants & Writing & .126 & .754 & 1.000\\
Main & Log named assistants & Disclosed & $< .001$ & .004 & .004\\
Main & Capped assistants & Writing & .096 & .649 & 1.000\\
Main & Capped assistants & Disclosed & $< .001$ & .002 & .002\\
Reference & Median reference age & Writing & .044 & .140 & .131\\
Reference & Median reference age & Disclosed & $< .001$ & $< .001$ & $< .001$\\
Reference & Log reference count & Writing & .338 & .487 & .565\\
Reference & Log reference count & Disclosed & .282 & .487 & .565\\
\bottomrule\end{tabular*}}\tabnote{Romano--Wolf step-down probabilities use 1,999 null-imposed wild-bootstrap draws, one Rademacher sign per researcher shared across models in each draw, and researcher/paper-clustered studentisation. Holm uses the registered twenty-test main and four-test reference families separately. These registered families are distinct from the exploratory decompositions displayed above, which sit outside the registered family. Neither adjustment covers source shocks.}\end{table}

\FloatBarrier\clearpage\setcounter{table}{0}\setcounter{figure}{0}
\section{Project choice and task savings}\label{app:model}
\enlargethispage{\baselineskip}
\subsection{Scope and team size}
Equation~\ref{eq:project} maximises cooperative surplus over integer $n\geq1$ and scope $x\geq0$. Assume $j>0$, $B(n)>0$ weakly increasing, $c_o'(a_o)<0$, and $K(n)$ growing sufficiently quickly to ensure a finite optimum. Credit allocation does not distort design; contributions may differ. The outside option is zero. Comparative statics hold other primitives fixed and condition on production.

At fixed $n$, strict concavity in scope gives the interior condition
\begin{equation}
 \frac{B(n)}{1+x^*}=c_o(a_o)+\frac{jx^*}{n},\qquad
 \frac{\partial x^*(n)}{\partial a_o}
 =\frac{-c_o'(a_o)}{B(n)/(1+x^*)^2+j/n}>0.
\end{equation}
Scope is zero when $B(n)\leq c_o$. For a smooth extension in team size,
\[
 U_{nx}=\frac{B'(n)}{1+x}+\frac{jx}{n^2}\geq0,
 \qquad U_{x,-c_o}=1,\qquad U_{n,-c_o}=0.
\]
The discrete condition follows directly: for $n'>n$, the derivative in $x$ of $U(n',x)-U(n,x)$ equals
\[
 \frac{B(n')-B(n)}{1+x}+jx\left(\frac1n-\frac1{n'}\right)\geq0.
\]
Scope and team size have increasing differences, as do scope and cheaper execution. The smallest and largest optimal choices therefore weakly increase as execution becomes cheaper, with discrete rosters changing at thresholds. This result places no sign restriction on comparisons between projects with different benefits, costs, or reporting practices.

\subsection{Writing savings and assistance hours}
The separable writing cost $C_w(a_w)$ cancels from comparisons between designs, leaving preferred scope and roster unchanged conditional on production. Lower costs can induce entry, changing average observed team size in either direction. Writing costs that depend on scope or the roster can instead change the preferred design. Separability is a maintained case, not a general property of writing assistance.

At an execution price normalised to one, write $e(a_o)=c_o(a_o)$. If all residual execution is delegated, assistant hours are $h=e(a_o)x$. Within a smooth choice region,
\begin{equation}
 \frac{dh}{da_o}=e'(a_o)x+e(a_o)\frac{dx}{da_o}.
\end{equation}
The first term is the saving on existing operations; the second is the work associated with expansion. Hours rise only when proportional scope expansion exceeds the proportional reduction in unit requirements. At roster thresholds the comparison is between optimised levels on either side. Translating hours into headcounts additionally requires hiring, allocation, and crediting rules. The number of acknowledged people does not supply those rules.

\subsection{Disclosure cutoff equilibrium}\label{sec:cutoff-equilibrium}
In the model of Section~\ref{sec:attribution-model}, a team discloses when $b$ exceeds the incremental reputational cost in equation~\ref{eq:attribution-payoff}. In a cutoff equilibrium with rational expectations, readers hold correct beliefs about $d_n$, and the cutoff $t_n$ and disclosure rate satisfy
\begin{equation}
 t_n=\frac{s(1-\pi)}{n(1-\pi d_n)},
 \qquad d_n=1-F(t_n),
 \qquad
 nt_n\{1-\pi+\pi F(t_n)\}=s(1-\pi).
 \label{eq:attribution-equilibrium}
\end{equation}
The last expression has a unique positive solution: its left-hand side is continuous and strictly increasing in $t_n$ from zero to infinity. At a fixed cutoff it also strictly increases with $n$, so the equilibrium cutoff falls as the author team grows. Positive density makes the disclosure rate interior and the comparison strict: $d_{n+1}>d_n$. For a continuous extension in $n$, implicit differentiation gives
\begin{equation}
 \frac{\mathrm d d_n}{\mathrm d n}
 =\frac{f(t_n)t_n\{1-\pi+\pi F(t_n)\}}
 {n\{1-\pi+\pi F(t_n)+\pi t_n f(t_n)\}}>0.
 \label{eq:attribution-comparative-static}
\end{equation}
The tie $b=t_n$ has measure zero under a continuous $F$. As $s\to0$ the cutoff tends to zero, every user discloses and the size gradient disappears. With $k$ exchangeable operators among $n$ authors, $1/n$ is replaced by $k/n$ in the payoff difference, and the gradient requires $k/n$ to fall with $n$. These derivations hold other primitives fixed; they do not map onto the fixed-effect regression coefficient, which conditions on researcher, source-quarter, and field effects.

\FloatBarrier\clearpage\setcounter{figure}{0}
\section{AI indicators by working-paper series}\label{app:series-trends}
\begin{figure}[!ht]\centering
\caption{Disclosed AI use by working-paper series}\label{fig:series-disclosed}
\includegraphics[width=\linewidth]{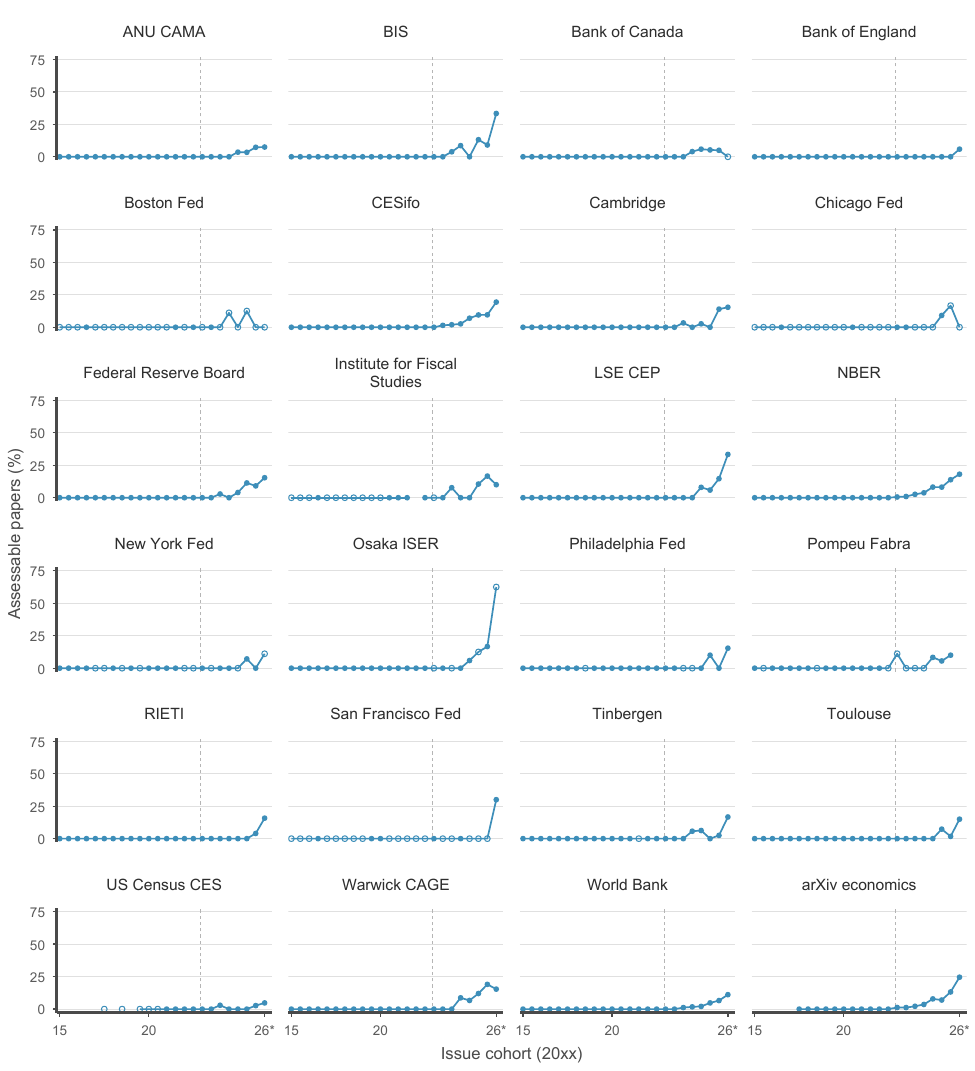}
\tabnote{Half-year issue cohorts, January 2015--October 2026; 26* denotes July--October 2026, an incomplete half-year. Each series uses its timing-supported papers with the indicator assessable, before regression restrictions. Axes are identical across panels and figures. Hollow points contain fewer than ten papers; missing cohorts are gaps. The dashed line marks November 2022. Printed-date observations retain the no-later-revision assumption described in Appendix~\ref{app:data}; cohort movements can reflect coverage and reporting changes.}
\end{figure}
\FloatBarrier
\clearpage
\begin{figure}[!ht]\centering
\caption{Writing flag by working-paper series}\label{fig:series-marker}
\includegraphics[width=\linewidth]{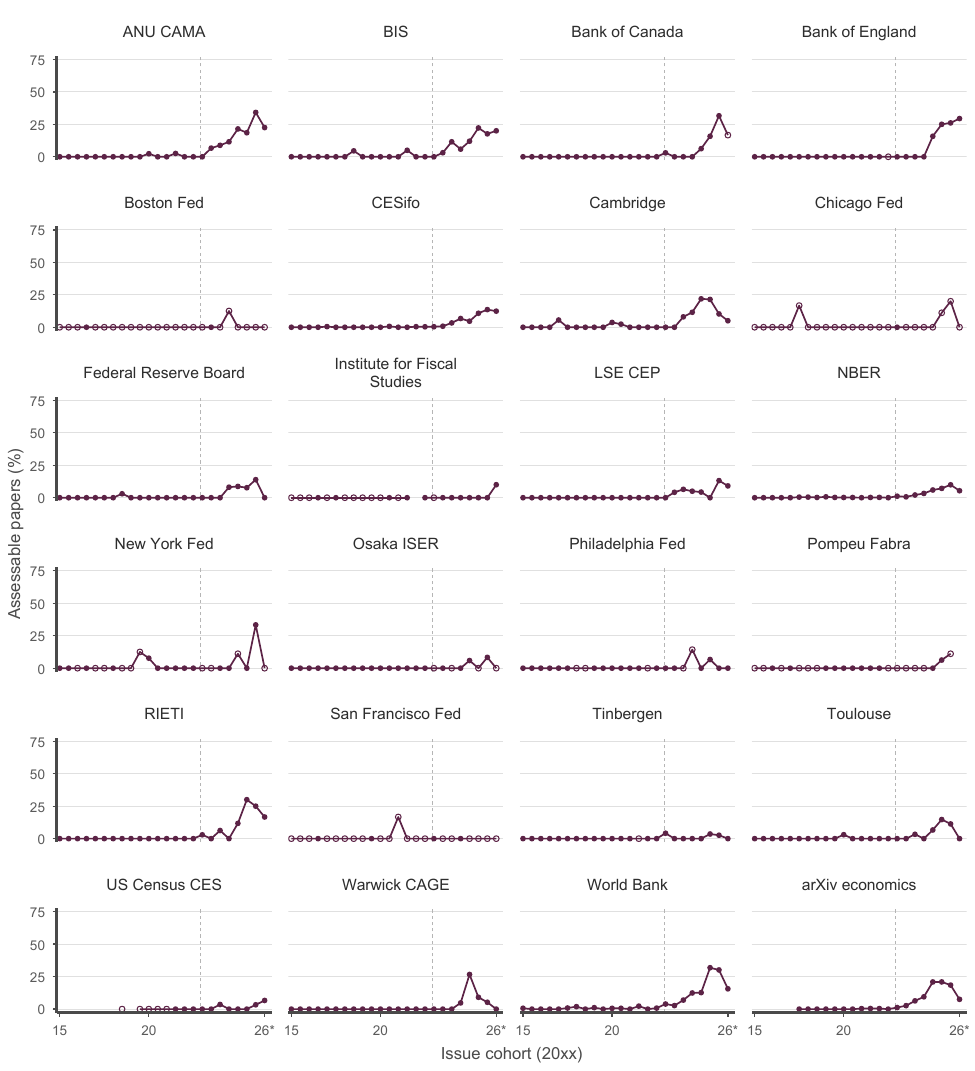}
\tabnote{Half-year issue cohorts, January 2015--October 2026; 26* denotes July--October 2026, an incomplete half-year. Each series uses its timing-supported papers with the indicator assessable, before regression restrictions. Axes are identical across panels and figures. Hollow points contain fewer than ten papers; missing cohorts are gaps. The dashed line marks November 2022. Printed-date observations retain the no-later-revision assumption described in Appendix~\ref{app:data}; cohort movements can reflect coverage and reporting changes.}
\end{figure}
\FloatBarrier

\end{document}